\documentclass[a4paper,fleqn]{cas-dc}

\usepackage[numbers]{natbib}
\usepackage{algorithmic}
\usepackage{algorithm}
\usepackage{graphicx}
\usepackage{epstopdf}
\usepackage{appendix}
\usepackage[section]{placeins}

\def\tsc#1{\csdef{#1}{\textsc{\lowercase{#1}}\xspace}}
\tsc{WGM}
\tsc{QE}
\tsc{EP}
\tsc{PMS}
\tsc{BEC}
\tsc{DE}
\begin{document}
\let\WriteBookmarks\relax
\def\floatpagepagefraction{1}
\def\textpagefraction{.001}
\shorttitle{Robust Convergency Indicator using MIMO-PI Controller in the presence of disturbances}
\shortauthors{ShengZimao et~al.}

\title [mode = title]{Robust Convergency Indicator using MIMO-PI Controller in the presence of disturbances} 
               
%\tnotemark[1,2]

\tnotetext[1]{This document is the results of the National Natural Science Foundation of China (Grant No.51775435), the pertinent preprint can be accessed via the following link \url{https://arxiv.org/abs/2411.13140}.}

\author[1,2]{Zimao Sheng}[style=chinese,
						type=editor,
                        auid=000,
                        %role=Researcher,
                        orcid=0000-0001-6067-3379
                        ]

%\fnmark[1]
%\ead{ShengZimao@mail.nwpu.edu.cn}
%\ead[url]{www.jkkrishnan.in}

\credit{Conceptualization of this study, Methodology, Software, Writing and Original draft preparation}

%\address[1]{, Street 129, 1043 NX Amsterdam, The Netherlands}
\affiliation[1]{organization={School of Mechanical Engineering, Northwestern Polytechnical University},
                addressline={127 West Youyi Road, Beilin District}, 
                city={Xi’an},
%               citysep={}, % Uncomment if no comma needed between city and postcode
                postcode={710072}, 
                state={Shaanxi},
                country={PR China}}

\author[1,2]{Hongan Yang}[style=chinese]
\cormark[1]
\ead{yhongan@nwpu.edu.cn}	
\credit{Provides idea guidance, some financial support and text revision opinions}

\author[1,2]{Jiakang Wang}
%[%
%   role=Co-ordinator,
%   ]
%\fnmark[2]
%\ead{wjh@example.org}
% \ead[URL]{https://www.university.org}

%\credit{Data curation, Writing - Original draft preparation}

\affiliation[2]{organization={Key Laboratory of Industrial Engineering and Intelligent Manufacturing},
     			addressline={127 West Youyi Road, Beilin District},
                postcode={710072}, 
                postcodesep={}, 
                city={Xi'an},
                state={Shaanxi},
                country={PR China}}
\credit{Discussion and part of the content writing}

\author[1,2]{Tong Zhang}
\credit{Discussion and part of the format typesetting}

%% 下面开始摘要部分
\begin{abstract}
	The PID controller remains the most widely adopted control architecture, with groundbreaking success across extensive implications. However, optimal parameter tuning for PID controller remains a critical challenge. Existing theories predominantly focus on linear time-invariant systems and Single-Input Single-Output (SISO) scenarios, leaving a research gap in addressing complex PID control problems for Multi-Input Multi-Output (MIMO) nonlinear systems with disturbances. This study enhances controller robustness by leveraging insights into the velocity form of nonlinear systems. It establishes a quantitative metric to evaluate the robustness of MIMO-PI controller, clarifies key theories on how robustness influences exponential error stabilization. Guided by these theories, an optimal robust MIMO-PI controller is developed without oversimplifying assumptions. Experimental results demonstrate that the controller achieves effective exponential stabilization and exhibits exceptional robustness under the guidance of the proposed robust indicator. Notably, the robust convergence indicator can also effectively assess comprehensive performance.
\end{abstract}
	

\maketitle

\section{Introduction}
\label{sec:Introduction}
\subsection{Motivation}
Classical proportional-integral-derivative (PID) control stands as the most fundamental and extensively utilized feedback - based control algorithm, being implemented in over 95$\%$ of control loops within engineering control systems \cite{bib:Åström}. Despite the advancements in advanced control techniques, PID control maintains its irreplaceable position \cite{bib:Samad}. This is attributed to its simple model-free structure and remarkable robustness in nullifying the impact of uncertainties. A plethora of enhanced PID controller versions \cite{bib:Blevins} have notably improved their key performance indicators, particularly in terms of robustness, in the domains of industrial process control \cite{bib:Kang} and nonlinear optimization problems \cite{bib:Alagoz}. However, most of the current industrial systems exhibit characteristics such as Multi-Inputs and Multi-Outputs (MIMO), strong coupling of state variables, strong external disturbances, undesired sensor faults and actuator faults. These features pose higher requirements for the design of a high-robustness PID controller with MIMO, nonlinearity, and anti-disturbance capabilities. 

Consequently, a pertinent question emerges: How to gauge the robustness of PID controller parameters for general perturbed MIMO nonlinear systems to enable more effective and optimal parameter regulation for exponential stabilization? 
Classical approaches to regulating PID controller parameters have predominantly relied on practical experiments \cite{bib:O'dwyer} and experiences like the Ziegler–Nichols rules \cite{bib:Ziegler}. For systems with specific structures, such as linear \cite{bib:Keel} or affine nonlinear \cite{bib:Killingsworth} systems, the gain regulation process for the PID controller is often manageable. 
Nevertheless, exploring suitable parameters for general nonlinear systems is usually complex as the corresponding search space is typically vast. 

Although there have been studies exploring methods to ensure the exponential stabilization of uncertain systems through appropriate PID parameter regulation \cite{bib:Zhao}\cite{bib:Zhao2}\cite{bib:Guo}, the current research emphasis still primarily lies in maintaining the asymptotic regulation of systems rather than enhancing the robustness capacity.
Besides, the control mechanism by which a PID controller can control MIMO nonlinear systems with disturbances remains unclear, impeding the implementation of a regulation method that can effectively guarantee robust stability.  
Moreover, the application of the theoretical results\cite{bib:Zhao} to practical parameter adjustment, and designing controllers applicable to MIMO nonlinear systems in a general form and with more general disturbances remains a formidable challenge. These circumstances have motivated us to consider the optimal parameter regulation of general perturbed MIMO systems to fortify the controller's robustness in the presence of disturbances.

\subsection{Related work}
The concept of addressing uncertainty using the infinity norm, known as $H_{\infty}$ control \cite{bib:Zames}, has ultimately been specialized to the case of robust PID control. In traditional robust PID control parameter tuning, frequency-domain internal model control (IMC) \cite{bib:Vilanova}\cite{bib:Rivera} is most commonly employed in linear Single-input and Single-output (SISO) systems. This is based on numerous variants of setting the gain and phase margin \cite{bib:Åström2}, as well as other flexible extensions that directly parameterize the maximum of the sensitivity function \cite{bib:Arrieta}. For nonlinear SISO systems, criteria such as the Popov criterion need to be considered when designing an appropriately robust stabilization utility \cite{bib:Son}. However, for a long time, a MIMO PID controller capable of stabilizing general nonlinear perturbed state-space-based systems has been absent. Additionally, manual PID parameter regulation undermines the controller's robustness, causing it to deviate from its optimal state and offering no guarantee of process stability \cite{bib:Verma}. Meanwhile, the explicit design formulas for PID parameters in the context of MIMO systems to globally stabilize the regulation error, along with theoretical insights relevant to a class of nonlinear, uncertain stochastic systems, are presented in \cite{bib:Zhao3}\cite{bib:Cong}\cite{bib:Zhang}. 

Theoretically, Cheng \cite{zhao2017pid} and Guo \cite{bib:Zhao} both derived sufficient conditions for ensuring effective error stabilization in PID controller. Specifically, controller parameters are constrained within a range linked to the Lipschitz constant of the nonlinear system's mapping. However, obtaining such a Lipschitz constant is typically challenging, which impedes practical implementation. Cheng \cite{bib:Zhao2} developed a PID controller by integrating high-order differential terms of the error. However, this focus often results in limited adaptability for stochastic systems affected by bounded perturbations. The underlying reason is the absence of customized robust indicators. Therefore, in this paper, a novel robustness criterion and an optimal fine-tuning for adaptive MIMO Proportional-Integral (MIMO-PI) controller parameter regulation methods are proposed, which ensures robust stability.\\
\subsection{Contributions}
The main contributions are as follows 
\begin{itemize}
	\item First, we innovatively established sufficient conditions for the robust stabilization of general disturbed autonomous nonlinear systems, alongside determining the specific exponential convergence rate and the range of the global random attractor. This forms a theoretical framework for subsequent quantification of key indicators of MIMO-PI controllers, as detailed in Theorem \ref{thm:lemma_A1};  
	\item Next, from the perspective of the velocity form, we analyzed the sufficient conditions for exponential convergence in disturbed nonlinear systems driven by MIMO-PI controllers (see Theorem \ref{thm:theorem_A4}). Using the controller coefficients, we quantified both the exponential convergence rate and the scope of the global random attractor, yielding the critical robust convergence indicators \( R_K \) and \( I_K \). These indicators were modeled as an eigenvalue problem (EVP), with specific calculation methods provided in Eq.(\ref{eq:R_K}) and Eq.(\ref{eq:I_K});
	\item Finally, building upon the aforementioned theory, we proposed a controller parameter optimization model that satisfies input constraints, as presented in Eq.(\ref{eq:optimal_model}). Experimental results validate the rationality of the indicators and the optimality of the optimization model.
\end{itemize}

This paper is organized as follows. After an overview in Section \ref{sec:Introduction}, the numerous theoretical results regarding the robustness of nonlinear perturbed systems intended for subsequent optimal MIMO-PI controller is given in Section \ref{sec:Preliminaries}, and the perturbed model and constrained input commands, along with the problem formulation, is presented in Section \ref{sec:Problem Formulation}. The proposed robust convergency indicator and optimal MIMO-PI controller are designed in Section \ref{sec:Controller Design for Optimizing Robust Stability}, and simulation verification experiments in Section \ref{sec:Simulation Verification}. Finally, we present concluding remarks and directions for future investigation in Conclusions and future work.

\section{Preliminaries}
\label{sec:Preliminaries}
In this section, we present several key definitions, lemmas, and theorem as essential tools for subsequent analyses.
\subsection{Notations and Definitions}
Denote $\mathbb{R}^n$ as the n-dimensional Euclidean space, $\mathbb{R}^{m\times n}$ as the space of $m\times n$ real matrices, $||x||_2 = x^Tx$ as the Euclidean 2-norm of a vector x. The norm of a matrix $P\in \mathbb{R}^{m\times n}$ is defined by $||P||_2=\sup_{x\in R^n,||x||_2=1}||Px||_2=\sqrt{\lambda_{\max}(P^TP)}$, for given matrix set $\mathbb{P}$, its 2-norm is defined as $||\mathbb{P}||_2=\arg\sup_{P\in\mathbb{P}}||P||_2$, $\sigma_{\min}(P)=\sqrt{\lambda_{\min}(P^TP)}$. We denote $Res(J)$ as the real part of the eigenvalues associated with matrix $J$, besides $\lambda_{\min}(S)$ and $\lambda_{\max}(S)$ as the smallest and the largest eigenvalues of $S$,respectively. For a function $\Phi =(\Phi_1,\Phi_2,...\Phi_n)^T \in \mathbb{R}^n$,$x=(x_1,x_2,...x_m)^T\in \mathbb{R}^m$, let $\frac{\partial \Phi}{\partial x^T} = (\frac{\partial \Phi_i(x)}{\partial x_j})_{ij}$. Matrix $A<B$ means $A-B$ is negative define matrix. 
\newtheorem{definition}{Definition}
\begin{definition}
	(Robust stability) For perturbed nonlinear system $\dot{y} = f(t,y) + g(t,y)$, $y \in \mathbb{R}^n$,$f(t,0)=0$, $f,g \in C[I\times S_H,\mathbb{R}^n]$,$S_H=\{x|||x||_2\leq H\}$, if $\forall \varepsilon >0$, there exists $\delta_1(\varepsilon) >0$ and $\delta_2(\varepsilon) >0$ to make $||g(t,y)||_2\leq \delta_1(\varepsilon)$, $||y(0)||_2\leq \delta_2(\varepsilon)$, and $||y||_2\leq \varepsilon$, then the trivial solution of $\dot{y} = f(t,y) + g(t,y),f(t,0)=0$ exhibits robust stability.
\end{definition}

\newtheorem{definition2}[definition]{Definition}
\begin{definition2}
	(Global attractor) For a dynamical system: $\dot{x}(t) = f(x(t))$, $x(t)\in \mathbb{R}^n$, where $f:\mathbb{R}^n\rightarrow \mathbb{R}^n$ is a deterministic vector field (typically Lipschitz continuous). The set $A$ for this state $x(t)$ is defined as the global attractor if $x(t)\notin A$, then the distance between $x(t)$ and $A$: $dist(x(t),A) \rightarrow 0$ as $t\rightarrow \infty$; if $x(t)\in A$, then $x(t+\Delta t) \in A$ for any $\Delta t \geq 0$.
\end{definition2}

\newtheorem{definition3}[definition]{Definition}
\begin{definition3}
	(Global random attractor) For a dynamical system governed by a stochastic differential equation (SDE): $\dot{x} = f(x) + d$, $x\in \mathbb{R}^n$ and $d$ is a stochastic perturbation. Its global attractor can be regarded as the global random attractor of the SDE.
\end{definition3}

%\newtheorem{definition4}[definition]{Definition}
%\begin{definition4}
%	(Co-Lipschitz property) For function $f:X\rightarrow \mathbb{R}^m$, $X\subset \mathbb{R}^n$, if there exists $L>0$ for any $x_1,x_2\in X$ such that $||f(x_1) - f(x_2)||_2\geq L||x_1-x_2||_2$ holds, then $f$ is referred to $L$-co-Lipschitz over $X$.
%\end{definition4}

\newtheorem{definition5}[definition]{Definition}
\begin{definition5}
	\label{def:smoothness}
	($\beta_f$-Smoothness)  A function $f:\mathbb{R}^n \rightarrow \mathbb{R}^n$ is called $\beta_f$-Smoothness if for any $x_1,x_2 \in \mathbb{R}^n$, there is
	\begin{equation}
	\| \frac{\partial f(x_1)}{\partial x} - \frac{\partial f(x_2)}{\partial x} \|_2 \leq \beta_f \|x_1-x_2\|_2
	\end{equation}
	where $\beta_f>0$, $\partial f(x_1)/\partial x$ and $\partial f(x_2)/\partial x$ are the Jacobians of $f$ at point $x_1$ and $x_2$ respectively.
\end{definition5}

\subsection{Key Lemmas and Theorem}
%\newtheorem{lem0}{Lemma}
%\begin{lem0}
%	\label{thm:lem0}
%	If there exists Lyapunov function $V(e)=e^TPe$, $P=P^T>0$ for automous system $\dot{e}=f(e)$ to have $\dot{V}(e)\leq -\alpha V(e)$, $\alpha >0$, then $e$ will be exponentially converge to 0 and we define its convergency rate is $-\alpha$.
%\end{lem0}
\newtheorem{lem0}{Lemma}
\begin{lem0}
	\label{thm:lem0}
	(Stability of matrix\cite{bib:Schaft}) If $A$ is stable, which means $Res(A)<0$ or eigenvalue $0$ corresponds to the single characteristic factor, there exists only one positive define $P=P^T=P(A,\delta)>0$ for any $\delta >0$ to make:
	\begin{equation}
		P(A,\delta)A + A^TP(A,\delta) + \delta I  = O
	\end{equation}
\end{lem0}
%\newtheorem{schur}[lem0]{Lemma }
%\begin{schur}
%	\label{thm:schur}
%	(Schur complement) For a given symmetric matrix $S=\begin{bmatrix}
%		S_{11} & S_{12} \\ S_{21} & S_{22}
%	\end{bmatrix}$, where $S_{11} \in \mathbb{R}^{r\times r}$. The following three conditions are equivalent. (i) $S<0$; (ii) $S_{11}<0,S_{22} - S_{12}^TS_{11}^{-1}S_{12}<0$; (iii) $S_{22}<0,S_{11} - S_{12}S_{22}^{-1}S_{12}^T<0$. 
%\end{schur}
\newtheorem{Wely}[lem0]{Lemma}
\begin{Wely}
	\label{lem:Wely}
	(Weyl's inequality\cite{horn2012matrix}) In the context of matrix perturbation analysis, consider a symmetric matrix \(A = A^T \in \mathbb{R}^{n \times n}\) and its perturbation \(\Delta A = \Delta A^T \in \mathbb{R}^{n \times n}\). A fundamental result states that\(\|\lambda_{\max}(A + \Delta A) - \lambda_{\max}(A)\|_2 \leq \|\Delta A\|_2,\)where \(\lambda_{\max}(\cdot)\) denotes the maximum eigenvalue. 
\end{Wely}

\newtheorem{lem1}[lem0]{Lemma}
\begin{lem1}
	\label{lem:lem4}
	For \(f(x) \in \mathbb{R}^n\), if \(f(x)\) is differentiable and \(\beta_f\)-Smoothness over the region \(x \in \Omega \subset \mathbb{R}^n\) near \(x = 0\), then there exists
	\begin{equation}
		\begin{split}
			&\left|
			\lambda_{\max}\left[ \frac{\partial (f(x) + f(x)^T)}{\partial x} \right] - \lambda_{\max}\left[ \frac{\partial (f(0) + f(0)^T)}{\partial x}\right] 
			\right| \\
			&\leq 2\beta_f ||x||_2
		\end{split}
	\end{equation}
\end{lem1}
\newproof{pf_lem1}{proof}
\begin{pf_lem1}
	There exists
	\begin{align}
		\footnotesize
		&\left|
		\lambda_{\max}\left[ \frac{\partial (f(x) + f(x)^T)}{\partial x} \right] - \lambda_{\max}\left[ \frac{\partial (f(0) + f(0)^T)}{\partial x}\right] 
		\right| \\
		&\leq
		||
		\frac{\partial (f(x) + f(x)^T)}{\partial x}
		-
		\frac{\partial (f(0) + f(0)^T)}{\partial x}
		||_2, (\mathbf{Lemma}\ \ref{lem:Wely}) \\
		& \leq
		2|| 
		\frac{\partial f(x) }{\partial x} 
		-
		\frac{\partial f(0) }{\partial x}
		||_2, (\beta_f-Smoothness\ of\ f(x) )\\
		& \leq 2\beta_f ||x||_2
	\end{align}
	The proof is completed.
\end{pf_lem1}

\newtheorem{lem2}[lem0]{Lemma}
\begin{lem2}
	\label{lem:lem2}
	Consider the function \(f(x) = A(x)x\), where \(x \in \mathbb{R}^n\), \(f(x) \in \mathbb{R}^n\), and \(A(x) \in \mathbb{R}^{n \times n}\). If \(f(x)\) is differentiable at \(x = 0\), the Jacobian at this point \( \partial f(0)/\partial x = A(0)\).
\end{lem2}
\newproof{pf_lem2}{proof}
\begin{pf_lem2}
	Let the $i$-th component of the function $f(x) = (f_1(x),f_2(x)\ldots f_n(x))^T$ is denoted as $f_i(x)$, $A(x)=[a_{ij}(x)]_{n\times n}$, the partial derivative of \(f_i(x)\) with respect to \(x_k\) is: 
	\begin{align}
		\frac{\partial f_i(x)}{\partial x_k} 
		&= \sum_{j=1}^{n}\left[
		\frac{\partial a_{ij}(x)}{\partial x_k} x_j + a_{ij}(x) \frac{\partial x_j}{\partial x_k}
		\right] \\
		&= \sum_{j=1}^{n}\frac{\partial a_{ij}(x)}{\partial x_k}x_j + a_{ik}(x)
	\end{align}
	Thus, for any $i,k$ there exists $\partial f_i(0)/\partial x_k = a_{ik}(0)$, the proof is completed.
\end{pf_lem2}

\newtheorem{lem3}[lem0]{Lemma}
\begin{lem3}
	\label{lem:lem3}
	For a nonlinear system $\dot{x}(t) =f(x(t))$ with initial state $x(0)=x_0,f(0)=0$, $x\in \mathbb{R}^n$, suppose there exists a positive definite function $V(x)$ for any nonzero $x$ satisfying
	\begin{equation}
		\label{equ:lem1_equ}
		\dot{V}(x) + \alpha V(x) - \beta V^{\frac{1}{2}}(x)  \leq 0
	\end{equation}
	where $\alpha >0$, $\beta >0$. Then for any state $x(t)$ there exists
	\begin{equation}
		\label{equ:lem2_equ}
		V^{\frac{1}{2}}(x(t)) \leq \frac{\beta}{\alpha} + (V^{\frac{1}{2}}(x_0) - \frac{\beta}{\alpha})e^{-\frac{\alpha}{2}t}
	\end{equation}
\end{lem3}
\newproof{pf_lem3}{proof}
\begin{pf_lem3}
	We define \(s(t) = V^{\frac{1}{2}}(x)\), and then Eq.(\ref{equ:lem1_equ}) can be transformed into
	\begin{equation}
		\dot{s}(t) + \frac{\alpha}{2}s(t) \leq \frac{\beta}{2}
	\end{equation}
	The above equation can be further transformed into 
	\begin{equation}
		e^{\frac{\alpha}{2}t}\dot{s}(t) + \frac{\alpha}{2}e^{\frac{\alpha}{2}t}s(t) \leq \frac{\beta}{2}e^{\frac{\alpha}{2}t}
	\end{equation}
	which means
	\begin{equation}
		\frac{d}{dt}(e^{\frac{\alpha}{2}t}s(t)) \leq \frac{\beta}{2} e^{\frac{\alpha}{2}t}
	\end{equation}
	Integrating both sides of the inequality in the above equation from $0$ to $t$ simultaneously, we can obtain
	\begin{equation}
		e^{\frac{\alpha}{2}t}s(t) - s(0) \leq \frac{\beta}{\alpha} (e^{\frac{\alpha}{2}t} - 1)
	\end{equation}
	Therefore, the above equation can be further derived to obtain Eq.(\ref{equ:lem2_equ}). Thus, the proof is completed.
\end{pf_lem3}
\newtheorem{rem0}{Remark}
\begin{rem0}
	\label{rem:rem0}
	Eq.(\ref{equ:lem2_equ}) reveals a fact that, when the initial value \(x_0\) satisfies \(V^{\frac{1}{2}}(x_0)\leq \frac{\beta}{\alpha}\),  \(V^{\frac{1}{2}}(x(t))\leq \frac{\beta}{\alpha}\) holds for all \(t\geq t_0\). Conversely, if \(V^{\frac{1}{2}}(x_0)\geq \frac{\beta}{\alpha}\), \(V^{\frac{1}{2}}(x(t))\) is upper - bounded by a function that converges exponentially to \(\frac{\beta}{\alpha}\) with a rate of \(\alpha\). As \(t\) approaches infinity, once \(V^{\frac{1}{2}}(t_r)\leq \frac{\beta}{\alpha}\) is satisfied at a certain time \(t_r\), the trajectory of \(x(t)\) will remain within this region indefinitely. 
	Consequently, it can be deduced that for any initial value $x(t_0) = x_0$, \(x(t)\) will ultimately converge to the region of attraction described as:
	\begin{equation}
		x(t) \in \{x|V^{\frac{1}{2}}(x)\leq \frac{\beta}{\alpha}\}, t\rightarrow \infty
	\end{equation}
\end{rem0}
It is worth noting that we generally aspire for the convergence rate \(\alpha\) to be as large as possible. Meanwhile, a larger \(\alpha\) also enables the scope of the final global random attractor, \(\frac{\beta}{\alpha}\), to be minimized. 

\newtheorem{thm0}{Theorem}
\begin{thm0}
	\label{thm:lemma_A1}
	For the perturbed autonomous system \(\dot{x} = f(x)+d\), where $x\in \mathbb{R}^n$,\(x(t_0)=x_0\), \(f(0)=0\), and \(d\in \mathbb{R}^n\) is the perturbation term satisfying \(\|d\|_2\leq L_d\). If there exists a  \(\Omega\) such that when \(x\in\Omega\), the following conditions hold: \\
	1) There exists \(L_f(\Omega)\geq0\) such that \(\left\|\frac{\partial f(x)}{\partial x}\right\|_{2}\leq L_f(\Omega)\);\\ 
	2) There exist a positive-definite matrix \(P = P^T > 0\) and \(\varepsilon(\Omega)>0\) such that \(\left[\frac{\partial f(x)}{\partial x}\right]^T P+P\frac{\partial f(x)}{\partial x}+\varepsilon(\Omega) I\leq O\).\\
	Then the ultimate trajectory of the state \(x(t)\), for any initial value $x_0$, \(x(t)\) will exponentially converge to the following global random attractor $S(\Omega)$ at least the rate of $\varepsilon(\Omega)/\lambda_{\max}(P)$:
	\begin{equation}
		\label{eq:lemma_A1_condition}
		x(t)\rightarrow S(\Omega) = \{x: ||f(x)||_2\leq \frac{2L_dL_f(\Omega)\lambda^2_{\max}(P)}{\varepsilon(\Omega) \lambda_{\min}(P)} \}, t\rightarrow \infty
	\end{equation}
	Simultaneously, for any \(x_0\notin S(\Omega)\), \(x(t)\) will converge exponentially to \(S(\Omega)\) at a rate of \(\varepsilon(\Omega)/\lambda_{\max}(P)\).
\end{thm0}
\newproof{pf_thm0}{proof}
\begin{pf_thm0}
	We construct the Lyapunov function 
	\begin{align}
		V(x)=f(x)^TPf(x)
	\end{align}
	such that
	\begin{equation}
		\begin{split}
			\dot{V}(x) = & \dot{f}(x)^TPf(x) + f(x)^TP\dot{f}(x) \\
			= & \dot{x}^T[\frac{\partial f(x)}{\partial x}]^TPf(x) + f(x)^TP[\frac{\partial f(x)}{\partial x}]\dot{x} \\
			= & f(x)^T([\frac{\partial f(x)}{\partial x}]^TP + P\frac{\partial f(x)}{\partial x})f(x)\\ 
			&+ d^T[\frac{\partial f(x)}{\partial x}]^TPf(x) 
			+  f(x)^TP\frac{\partial f(x)}{\partial x} d \\
			\leq & -\frac{\varepsilon(\Omega)}{\lambda_{\max}(P)}V(x) + 2d^T[\frac{\partial f(x)}{\partial x}]^TPf(x) \\
			\leq & -\frac{\varepsilon(\Omega)}{\lambda_{\max}(P)}V(x) + 2L_d||\frac{\partial f(x)}{\partial x}||_2||Pf(x)||_2\\
			\leq& -\frac{\varepsilon(\Omega)}{\lambda_{\max}(P)}V(x) + \frac{2L_dL_f(\Omega) \lambda_{\max}(P)}{\sqrt{\lambda_{\min}(P)}}V(x)^{\frac{1}{2}}
		\end{split}
	\end{equation}   
	According to the Lemma \ref{lem:lem3} and Remark \ref{rem:rem0}, as $t\rightarrow \infty$, 
	\begin{equation}
		V^{\frac{1}{2}}(x) \leq \frac{2L_dL_f(\Omega)\lambda^2_{\max}(P)}{\varepsilon(\Omega) \sqrt{\lambda_{\min}(P)}}
	\end{equation}
	Consequently, as $t\rightarrow \infty$,
	\begin{equation}
		||f(x)||_2\leq\frac{V^{\frac{1}{2}}(x)}{\sqrt{\lambda_{\min}(P)}}\leq \frac{2L_dL_f(\Omega)\lambda^2_{\max}(P)}{\varepsilon(\Omega) \lambda_{\min}(P)}
	\end{equation}
	The proof is completed.
\end{pf_thm0}
Theorem \ref{thm:lemma_A1} elucidates a pivotal theoretical framework. Let \(\Omega\) be a non-empty subset of the state space \(\mathbb{R}^n\) that encompasses the origin. When the Jacobian of the function \(f(x)\), denoted as \(J_f(x) = \partial f(x)/\partial x\), is uniformly bounded for all \(x\) within the domain \(\Omega\), and \(J_f(x)\) exhibits negative definiteness throughout this region, the following property emerges for the associated dynamical system. For an autonomous dynamical system governed by the differential equation \(\dot{x}(t)=f(x(t))+d\), regardless of the chosen initial condition \(x_0\in\mathbb{R}^n\), in the absence of external perturbations, the time-derivative of the state trajectory \(f(x(t))\) converges exponentially to the region of attraction as delineated in Eq.(\ref{eq:lemma_A1_condition}). The spatial extent of this region of attraction is intricately linked to two critical parameters: the supremum \(L_f(\Omega)\) of the norm of the Jacobian matrix \(J_f(x)\) over the domain \(\Omega\), and the upper bound \(-\varepsilon(\Omega)\) of the maximum eigenvalue of \(J_f(x)\) within \(\Omega\). Specifically, a smaller upper bound for the ultimate global random attractor \(\|f(x)\|_2\) can be achieved through a smaller \(L_f(\Omega)\) and a larger \(\varepsilon(\Omega)\).
\newtheorem{rem1}{Remark}
\begin{rem1}
	\label{rem:rem_lemma_A1}
	According to Lemma \ref{thm:lem0}, there exists a sufficient condition for the validity of condition 2) in Theorem \ref{thm:lemma_A1}, specifically: when \(x \in \Omega\), the Jacobian \( \partial f/\partial x \) is stable. Furthermore, a special case arises when \(\Omega = \{0\}\). In this scenario, determining whether \(\partial f(0)/\partial x\) satisfies conditions 1) and 2) suffices to characterize the global random attractor \(S(0)\) of the perturbed system's trajectory.
\end{rem1}
Remark \ref{rem:rem_lemma_A1} establishes the following theoretical principle: if the Jacobian matrix \( \partial f/\partial x \) of \(f(x)\) is continuously bounded and stable over a region \(x \in \Omega\), then the trajectory \(x(t)\) of the perturbed system will exponentially converge to a global random attractor in which the fluctuation amplitude of \(f(x)\) remains bounded. Notably, the extent of this global random attractor is governed by the eigenvalues of \( \partial f/\partial x \)—larger eigenvalue magnitudes generally correspond to a smaller constrained domain, thereby enabling the system to attain a higher level of robust stability.

\section{Problem Formulation}
\label{sec:Problem Formulation}
In this section, based on Section \ref{sec:Preliminaries}, we analyze a general perturbed nonlinear system and propose a optimal MIMO-PI controller for optimizing its robustness near the origin. 
This method transforms the robustness optimization problem on a global scale into a special type of eigenvalue optimization problem at the origin, which can then be efficiently solved using classical methods. 
%\subsection{The perturbed control system}

Consider the following perturbed autonomous MIMO non-affine nonlinear system within continuous and first-order differentiable $f \in \mathbb{R}^n$ and first-order differentiable disturbance $d \in \mathbb{R}^n$
\begin{align}
	\label{eq:perturbed non-affine nonlinear system}
	\begin{cases}
		\dot{x}(t) = f(x(t),u(t)) + d \\
		y(t) = h(x(t))\\
		x(0)=x_0, f(0)=0\\
	\end{cases}
\end{align}
where state $x(t)\in \mathbb{R}^n$, control input $u(t) \in \mathbb{R}^{m}$, observable $y(t)\in \mathbb{R}^L$. The disturbance term \(d \in \mathbb{R}^n\) has an upper bound on its amplitude as follows
\begin{equation}
	||\dot{d}||_2 \leq L_d
\end{equation} 
It is recognized that, when random disturbances are present, errors in the vicinity of the origin are ultimately inescapable. 

Nevertheless, our objective is to devise a suitable controller that, upon its application for any initial state $x_0$ and region $\Omega$, minimizes the terminal-state error in the neighborhood of the origin, which aims to make the controlled variable $x(t)$ converge to the vicinity of desired reference value $x_r\in \mathbb{R}^n$ exponentially to the nearest extent possible. Specifically: 
\begin{equation}
	\label{eq:goal}
	\min_{u} ||x(t)||_2,\  x(t) \in \Omega,\quad t\rightarrow \infty
\end{equation}

In this context, we assume the existence of a state observer \(h^{-1}:\mathbb{R}^{m} \to \mathbb{R}^{n}\), which is capable of delivering an exact state estimate \(x(t) = h^{-1}(y(t))\) based on the current observable \(y(t)\). This observer ensures that for every value of the observable quantity \(y(t)\), the estimated state \(x(t)\) precisely reflects the actual state of the system, thereby bridging the gap between the measured output and the internal states.

Further, we assume the adoption of a MIMO-PI controller that takes into account the coupling of multiple input channels. Explicitly, 
\begin{equation}
	\label{eq:controller}
	u(t) = K_P x(t) + K_I \int_0^t x(t) dt
\end{equation}
where $K_P \in \mathbb{R}^{m \times n}, K_I \in \mathbb{R}^{m \times n}$ and $ x(t) = h^{-1}(y(t))$.
Concurrently, the robustness optimization objective of Eq.(\ref{eq:goal}), can be recast as an optimization task focused on determining the optimal values of the weights \(K_P, K_I\). In what follows, we will elucidate the approach of leveraging the solution of a specific class of EVP to accomplish this goal. Here, the control commands and its first-order derivative quantity are constrained as
\begin{equation}
	\label{equ:commands_constrain}
	u_{\min}\leq u(t) \leq u_{\max},  \dot{u}_{\min} \leq \dot{u}(t) \leq \dot{u}_{\max}
\end{equation}
where $u_{\min},u_{\max},\dot{u}_{\min},\dot{u}_{\max},\in \mathbb{R}^{m\times 1}$. We expect to exponentially stablize the Eq.(\ref{eq:perturbed non-affine nonlinear system}) within maximum robust indicator to resist the emergency situation. 

\section{Controller for Optimizing Robust Stability}
\label{sec:Controller Design for Optimizing Robust Stability}
Our objective is to minimize the ultimate value of the norm $\|x(t)\|_2$ for the controller designed based on Eq. (\ref{eq:controller}). This goal corresponds to confining the state trajectory $x(t)$ to fluctuate within a narrow neighborhood around the origin. Leveraging the insights from Theorem \ref{thm:lemma_A1}, we are able to derive the sufficient conditions to meet our objective and solve it via classical optimization algorithms.
\subsection{Theoretical Foundation}
\newtheorem{thm1}[thm0]{Theorem}
The core theorems of this paper are presented below, which serve to characterize the relationship between the robust stability of the MIMO-PI controller and its coefficients.
\begin{thm1}
	\label{thm:theorem_A4}
	For the perturbed autonomous system \(\dot{x}(t) = f(x(t),u(t))+d\), where $x(t)\in \mathbb{R}^n$,\(x(t_0)=x_0\), \(f(0)=0\), and \(d\in \mathbb{R}^n\) is the perturbation term satisfying \(\|\dot{d}\|_2\leq L_d\), the MIMO-PI controller $u(x(t))=K_P x(t) + K_I \int_0^t x(t) dt$ ,  $K_P \in \mathbb{R}^{m \times n}, K_I \in \mathbb{R}^{m \times n}$ is adopted as the input. 
	If the following conditions are satisfied:\\
	1) Here $f(x) = f(x,u(x))$ is differentiable at \(x = 0\), there exists $L_f > 0$ such that 
	\begin{align}
		\left\| \frac{\partial f(0)}{\partial x} \right\|_2
		\leq L_f
	\end{align}
	and $f(x)$ is $\beta_f$-Smoothness near the origin; \\
	2) Let
	\begin{align}
		A_K(0) = \begin{pmatrix}
			\frac{\partial f(0)}{\partial x} + \frac{\partial f(0)}{\partial u}K_P & \frac{\partial f(0)}{\partial u}K_I \\
			I_n & O
		\end{pmatrix}
	\end{align}
	Suppose appropriate values of \(K_P, K_I\) are chosen such that
	\begin{align}
		Re\left[A_K(0)\right]<0
	\end{align}
	Then, the ultimate trajectory of the state \(x(t)\), as $t\rightarrow \infty$, must converge exponentially to
	\begin{align}
		\label{eq:x_upperbound_}
		s(t) = (x(t)^T,\dot{x}(t)^T)^T \rightarrow
		B(0,2L_dL_fI_K)	 
	\end{align}
	with at least the rate of \(R_K\), where $B(0,r)=\{x:||x||_2\leq r\}$, 
	\begin{align}
		\label{eq:robustness_indictor}
		R_K = \frac{\varepsilon(0)}{\lambda_{\max}(P)},\  I_K =\frac{||A_K(0)^{-1}||_2\lambda_{\max}^2(P)}{\varepsilon(0)\lambda_{\min}(P)} 
	\end{align}
	and $P$ is positive define matrix satisfying $P=P^T>0$ and
	\begin{align}
		\label{eq:A_K}
		A_K(0)^TP+PA_K(0) + \varepsilon(0)I\leq O
	\end{align}
\end{thm1}
\newproof{pf_thm1}{proof}
\begin{pf_thm1}
	Substituting the differential form of the MIMO-PI controller
	\begin{align}
		\dot{u}(t)=K_P \dot{x}(t) + K_I x(t)
	\end{align} into the velocity form of  perturbed autonomous system\cite{WOS:000073773400007}
	\begin{align}
		\label{eq:velocity_form}
		\ddot{x}(t) = \frac{\partial f}{\partial x}\dot{x}(t) + \frac{\partial f}{\partial u}\dot{u}(t)+\dot{d}
	\end{align}
	to obtain the state-space model of $s(t)=(\dot{x}(t),x(t))^T$ considering the perturbed term $d_s = (\dot{d}^T,0^T)^T$ as
	\begin{align}
		\begin{pmatrix}
			\ddot{x}(t) \\ \dot{x}(t)
		\end{pmatrix}
		&=
		\begin{pmatrix}
			\frac{\partial f}{\partial x} + \frac{\partial f}{\partial u}K_P & \frac{\partial f}{\partial u} K_I\\
			I_n & O
		\end{pmatrix}
		\begin{pmatrix}
			\dot{x}(t) \\ x(t)
		\end{pmatrix}
		+
		\begin{pmatrix}
			\dot{d} \\ 0
		\end{pmatrix}
		\\
		&\iff
		\dot{s}(t) = f_s(s(t)) + d_s = A_K(s(t)) s(t) + d_s
	\end{align}
	Furthermore, we can simplify the Jacobian of $f_s(0)$ according to the Lemma \ref{lem:lem2} that 
	\begin{align}
		\frac{\partial f_s(0)}{\partial x} = A_K(0) \ is \ stable
	\end{align}
	Hence, according to Lemma \ref{thm:lem0}, there exists $P=P^T>0$ and $\varepsilon(0) > 0$ such that
	\begin{align}
		\left[ \frac{\partial f_s(0)}{\partial x}\right]^TP + P \left[ \frac{\partial f_s(0)}{\partial x} \right] + \varepsilon(0) I \leq O
	\end{align}
	Let $\Omega = \{0\}$, according to Theorem \ref{thm:lemma_A1} and Remark \ref{rem:rem_lemma_A1}, as $t\rightarrow \infty$
		\begin{equation}
		\label{eq:lemma_A3_condition}
		s(t)\rightarrow S(0) = \left\{ x: ||A_K(0)x||_2\leq \frac{2L_dL_f\lambda^2_{\max}(P)}{\varepsilon(0) \lambda_{\min}(P)} \right\} 
	\end{equation}
	Given that $Re[A_K(0)] < 0$, it follows that $A_K(0)$ is non-singular, and consequently, there exists a set 
	\begin{align}
		S^* = \left\{ x: ||x||_2\leq \frac{2L_dL_f||A_K(0)^{-1}||_2\lambda^2_{\max}(P)}{\varepsilon(0) \lambda_{\min}(P)} \right\} 
	\end{align}
	such that $s(t) \rightarrow S(0) \subset S^*$. Thus, the proof is completed.
\end{pf_thm1}
\newtheorem{rem2}[rem1]{Remark}
\begin{rem2}
It is worth noting that when \( L_d \rightarrow 0 \), the system's final state \( s(t) \) will converge exponentially to zero, which precisely demonstrates its robust stability. Meanwhile, this also indirectly verifies that the MIMO-PI controller can effectively stabilize the linearized system near the equilibrium point, thereby achieving global exponential stabilization for the corresponding nonlinear system.
\end{rem2}
Meanwhile, it also should be noted that $A_K(0)$ can be described as a linear combination of the coefficients $K_P$ and $K_I$ as follows
\begin{align}
	A_K(0) = D_1(0) + D_2(0) K  
\end{align}
where 
\begin{align}
	K = (K_P,K_I)
\end{align}
and
\begin{align}
	D_1(0) = \begin{pmatrix}
		\frac{\partial f(0)}{\partial x} & O\\
		I_n & O
	\end{pmatrix},\ 
	D_2(0) = \begin{pmatrix}
		\frac{\partial f(0)}{\partial u} \\ O
	\end{pmatrix}\
\end{align}
\subsection{Robust Convergency Indicator}
From the perspective of Theorem \ref{thm:theorem_A4}, two indices \( R_K \) and \( I_K \) are introduced to characterize the convergence performance of the MIMO-PI controller. Specifically, \( R_K \) quantifies the exponential convergence rate of the error, while \( I_K \) delineates the range of the ultimate convergent global random attractor of the error. The relationship between \( I_K \) and \( R_K \) is characterized by
\begin{align}
	I_K=\frac{\tau(P_K)}{R_K\sigma_{\min}(A_K(0))}
\end{align} 
where
\begin{align}
	\tau (P_K) = \frac{\lambda_{\max}(P_K)}{\lambda_{\min}(P_K)},\
	R_K = \frac{\varepsilon(0)}{\lambda_{\max}(P_K)}
\end{align}
and subject to 
\begin{align}
	\label{eq:constrain}
 	A_K(0)^TP_K+P_KA_K(0) + \varepsilon(0)I\leq O
\end{align}
Furthermore, a larger \( R_K \) implies a faster error convergence rate, which is prone to inducing oscillations. Counterintuitively, an excessively large \( R_K \) does not necessarily guarantee that the system error will eventually converge to a smaller global random attractor. The ultimate global random attractor range \( I_K \) of the error during the convergence process is influenced by three factors: the condition number \( \tau(P_K) \) of the positive definite matrix \( P_K \), the exponential convergence rate \( R_K \), and the minimum spectral norm \( \sigma_{\min}(A_K(0)) \) of \( A_K(0) \).
\subsection{The Calculation of Indicator}
We aim to compute the corresponding \( R_K \) and \( I_K \) values for any given \( K_P \) and \( K_I \), which would allow us to screen out the optimal coefficients \( K_P \) and \( K_I \) for the MIMO-PI controller in any perturbed nonlinear system. The key point, and indeed the first step, of this problem is how to calculate \( R_K \). Since it is coupled with both the positive definite matrix \( P_K \) and the margin \( \varepsilon(0) \), computing its exact value poses a challenge. To tackle this issue, we introduce the positive definite matrix \( Q=Q^T>0 \) associated with \( P \) as follows, which serves as the starting point for computing \( R_K \).
\begin{align}
	Q_K= \frac{P_K}{\varepsilon(0)}
\end{align}
Thus, we can present the relationship between \( R_K \) and \( Q_K \) as follows:
\begin{align}
	R_K = \frac{1}{\lambda_{\max}(Q_K)}
\end{align}
subject to 
\begin{align}
	\label{eq:Q_set}
	A_K(0)^T Q_K + Q_K A_K(0) + I \leq O
\end{align}
In fact, there are theoretically numerous, even infinitely many, such \( Q_K \). We thus adopt an optimistic estimation: specifically, we take the value of \( R_K^* \) as our actual \( R_K \), where \( R_K^* \) corresponds to \( Q_K^* \)—the element that maximizes \( R_K \) within the set defined by Eq. (\ref{eq:Q_set}). Under this framework, the solution for \( Q_K^* \) is reduced to a type of eigenvalue problem (EVP) as follows:
\begin{align}
	& Q_K^*= \text{argmin}_{Q_K=Q_K^T > 0} \gamma \\
	& s.t.\ 
	\begin{cases}
		Q_K \leq \gamma I \\
		A_K(0)^T Q_K + Q_K A_K(0) + I \leq O	
	\end{cases}
\end{align}
Upon obtaining \( Q_K^* \), we can then compute \( R_K \) and \( I_K \) as follows:
\begin{align}
	\label{eq:R_K}
	R_K &= \frac{1}{\lambda_{\max}(Q_K^*)} \\
	\label{eq:I_K}
	I_K &= \frac{\tau(Q_K^*)}{R_K\sigma_{\min}(A_K(0))}
\end{align}
\subsection{Optimization Model of Controller Coefficients}
Our primary objective is to determine the optimal gain coefficients for the MIMO-PI controller. Specifically, under the constraints of a predefined exponential convergence rate \( R^* \), a given initial value \( x(t_0) = x_0 \), and input limitations as specified in Eq.(\ref{equ:commands_constrain}), we aim to maximize the achievement of the optimization objective outlined in Eq.(\ref{eq:goal}). This optimization model can be characterized by the indices \( R_K \) and \( I_K \) as follows
\begin{align}
	\label{eq:optimal_model}
	\begin{split}
		& \max_K R_K \\
		& s.t.\ 
		\begin{cases}
			I_K \leq I^* \\
			u_{\min} \leq u_K(t_0) \leq u_{\max} \\
			\dot{u}_{\min} \leq \dot{u}_K(t_0) \leq \dot{u}_{\max}
		\end{cases}		
	\end{split}	
\end{align}
where let sampling time $\Delta t = 0.1s$, and
\begin{align}
	u_K(t_0) = K \begin{pmatrix}
		x_0\\ x_0 \Delta t
	\end{pmatrix},\ 
	\dot{u}_K(t_0) = K \begin{pmatrix}
		0\\ x_0
	\end{pmatrix}
\end{align}
The aforementioned optimization model represents a canonical constrained nonlinear optimization problem (NP), which can be effectively addressed using algorithms such as the classical genetic algorithm (GA) to determine the optimal controller coefficients. Solving this model solely requires knowledge of the initial state of the perturbed system and the Jacobian matrix evaluated at the system's equilibrium point; explicit information regarding the disturbance type or the detailed system structure is not necessary. Furthermore, this approach obviates the need for extensive model simplification, a distinct advantage that facilitates its implementation in model-free engineering applications.

\section{Simulation Verification}
\label{sec:Simulation Verification}
In this section, we first verify the correctness of our proposed theoretical framework, specifically Theorem \ref{thm:lemma_A1}. Building upon this, we further validate the rationality of the robust convergence indicator proposed for the MIMO-PI controller through comparative experiments with different controller parameters.
\subsection{The Correctness of the Theorem \ref{thm:lemma_A1}}
\begin{figure*}[htbp]
	\centering
	\begin{tabular}{cc}
		\includegraphics[width=0.43\textwidth]{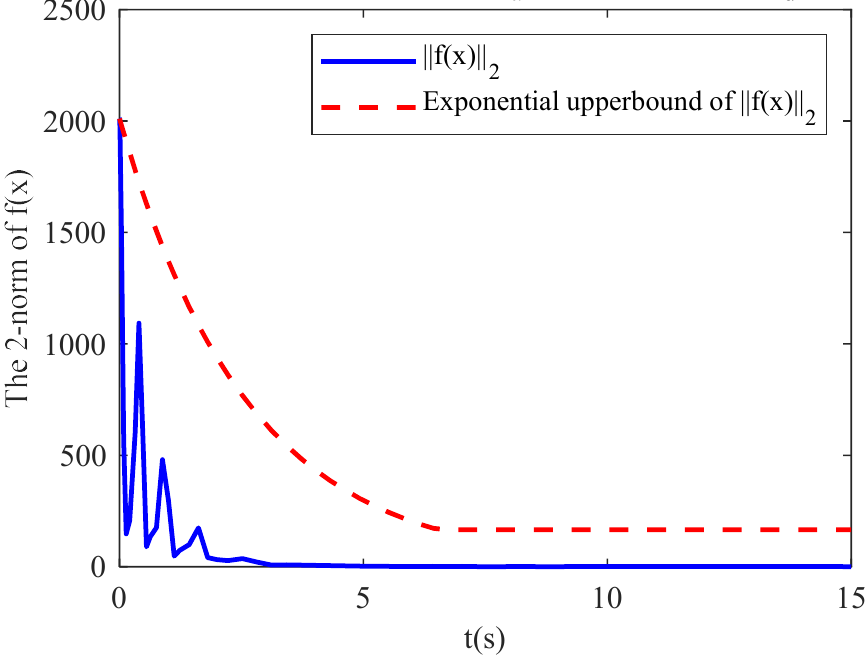} & 
		\includegraphics[width=0.43\textwidth]{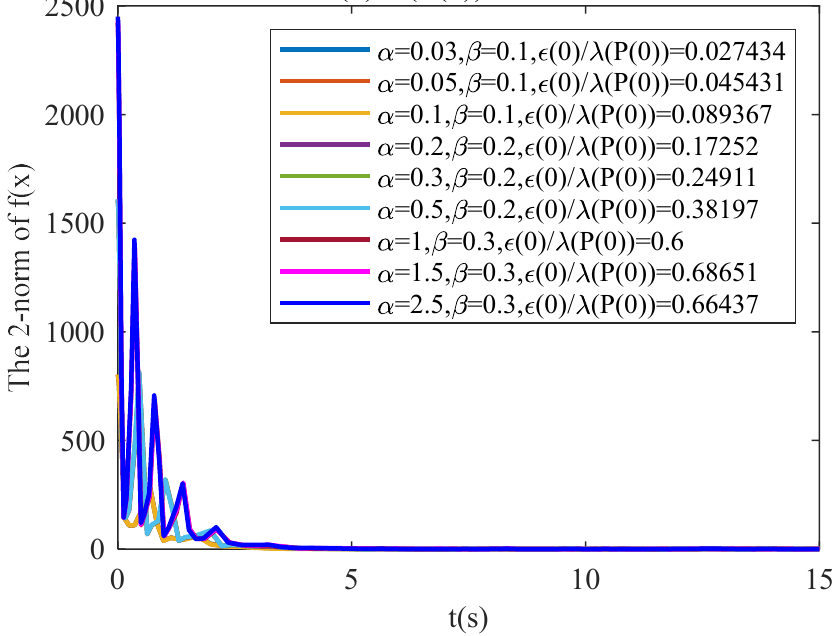} \\ 
		(a)& (b)
	\end{tabular}
	\begin{tabular}{cc}
		\includegraphics[width=0.43\textwidth]{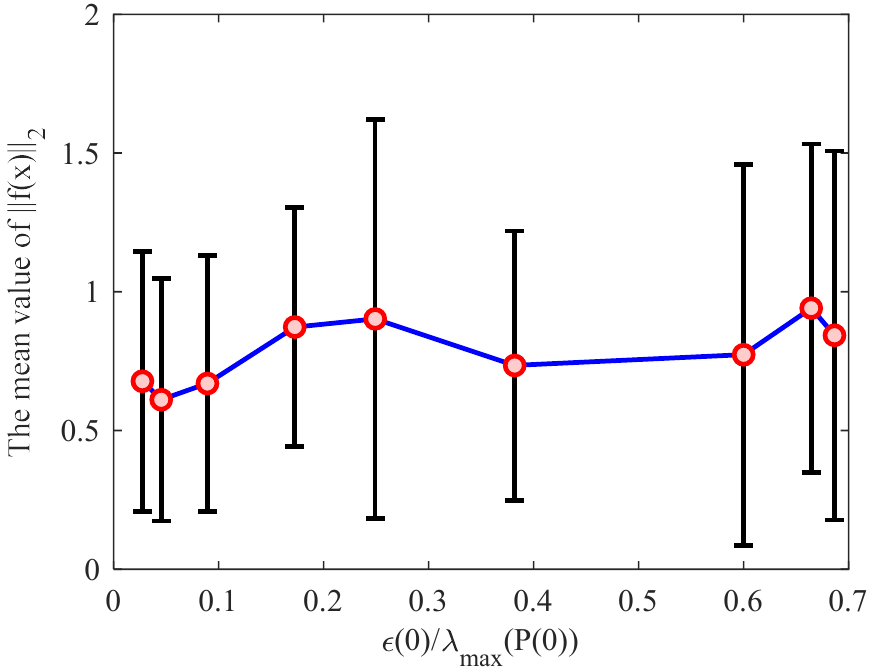} & 
		\includegraphics[width=0.43\textwidth]{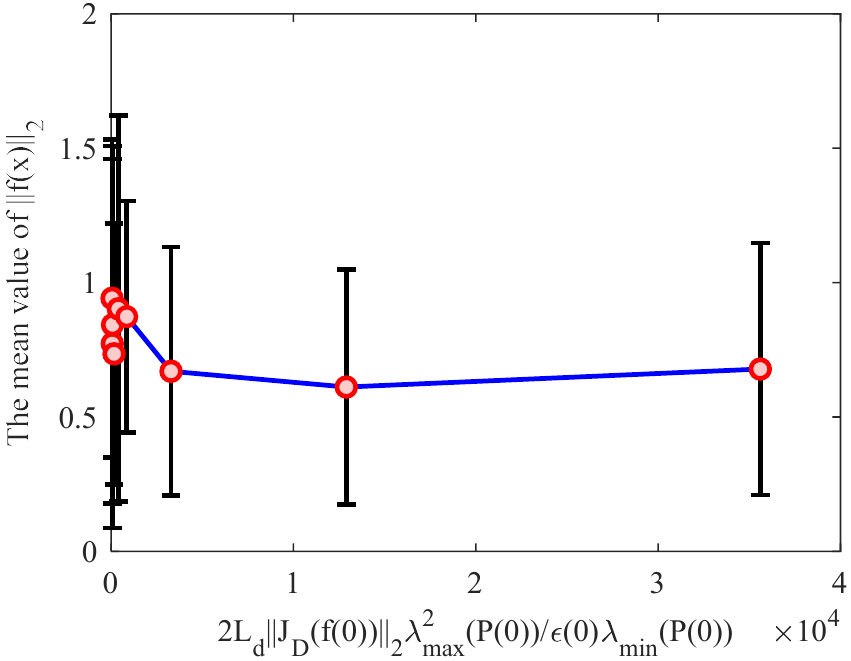} \\
		(c) & (d)
	\end{tabular}
	\caption{For multiple sets of simulation experiments on the Duffing model. (a): Curves of $\|f(x)\|_2$ and its exponential upperbound versus time for $\alpha = 0.5$, $\beta = 0.25$, $\delta = 1.5$, $L_d = 2.5$; (b): Time-series curves of $\frac{\varepsilon(0)}{\lambda_{\max}(P(0))}$ and $\|f(x)\|_2$ for different $\alpha$ and $\beta$; (c): Curves of the mean and standard deviation of $\|f(x)\|_2$ for different $\frac{\varepsilon(0)}{\lambda_{\max}(P(0))}$; (d): Curves of the mean and standard deviation of $\|f(x)\|_2$ for different $\frac{2L_d||J_D(f(0))||_2\lambda^2_{\max}(P(0))}{\varepsilon(0) \lambda_{\min}(P(0))}$.}
	\label{Fig:3_}
\end{figure*}

We will verify the validity of Theorem \ref{thm:lemma_A1} for classical perturbed nonlinear autonomous system in this subsection, such as Duffing model. The Duffing model\cite{WOS:000758238000013} is a nonlinear vibration model proposed by the German physicist Rudolf Duffing in the early 20th century. This model describes the vibration behavior of systems with nonlinear restoring forces and finds extensive applications in physics, engineering, biology, chaos theory, and other fields, particularly for studying bifurcation, chaos, and resonance characteristics in nonlinear systems. The Duffing model without external driving force and subject to bounded perturbation \(d_F\) can be described as:
\begin{equation}
	\ddot{x} + \delta \dot{x} + \alpha x + \beta x^3 = d_F, \ ||d_F||_2\leq L_d
\end{equation}
where \(x\) denotes the deformation of the damped nonlinear oscillator, while \(\alpha\), \(\beta\), and \(\delta\) represent the corresponding structural parameters. The corresponding perturbed nonlinear state-space model is described as follows:
\begin{equation}
	\label{eq:Duffing}
	\begin{cases}
		\dot{x}_1 = x_2 \\
		\dot{x}_2 = -\delta x_2 - \alpha x_1 - \beta x_1^3 + d_F
	\end{cases}
\end{equation}
The Jacobian of the above unperturbed system can be expressed as:
\begin{equation}
	\label{eq:Jacobian_Duffing}
	J_D(f) = 
	\begin{pmatrix}
		0 & 1 \\
		-\alpha-3\beta x_1^2 & -\delta
	\end{pmatrix}
\end{equation}
According to Theorem \ref{thm:lemma_A1} and Remark \ref{rem:rem_lemma_A1}, we can configure appropriate $\alpha$, $\beta$, and $\delta$ such that when $(x_1, x_2) = (0, 0)$, the Jacobian 
\begin{equation}
	J_D(f(0)) = 
	\begin{pmatrix}
		0 & 1 \\
		-\alpha & -\delta
	\end{pmatrix}
\end{equation}
satisfies the following condition that, there exist a positive-definite matrix \(P(0) = P(0)^T > 0\) and \(\varepsilon(0)>0\) such that
\begin{equation}
	\label{eq:Duffing_condition}
	J_D(f(0))^T P(0)+P(0)J_D(f(0))+\varepsilon(0) I\leq O
\end{equation}
A sufficient condition is:
\begin{equation}
	\lambda_{\max}(J_D(f(0))) < 0  \iff \alpha >0,\delta >0
\end{equation}
Then as $t\rightarrow \infty$, $x(t)$ would converge exponentially to
\begin{align}
	S(0) = \{x: ||f(x)||_2\leq \frac{2L_d||J_D(f(0))||_2\lambda^2_{\max}(P(0))}{\varepsilon(0) \lambda_{\min}(P(0))} \}
\end{align}
at least the rate of $\varepsilon(0)/\lambda_{\max}(P(0))$. Next, we will conduct multiple sets of comparative experiments to verify this point. As illustrated in Figure \ref{Fig:3_}, subfigure (a) validates the robust stability of the autonomous system under disturbances. Subfigure (b) compares convergence performance across different parameters \( \alpha \) and exponential convergence rates. Subfigures (c) and (d) quantitatively characterize the results from subfigure (b), with (c) and (d) respectively illustrating how \( R(f) \) and \( I(f) \) influence the mean and standard deviation of \( \| f(x) \|_2 \).  

\subsection{The Optimizable MIMO-PI Controller}
To validate the designed optimizable robust controller, its feasibility is verified herein through the simplified kinematic model\cite{WOS:001375965402029} of a fixed-wing aircraft in the ground coordinate system along the $\gamma$ and $\chi$ directions as follows
\begin{align}
	\label{eq:aircraft guidance law}
	\begin{split}
		&\dot{\chi}(t) = \frac{g\tan \phi(t)}{V} + d_{\chi} \\
		&\dot{\gamma}(t) = \frac{g(n_z(t)\cos \phi(t) - \cos \gamma (t))}{V} + d_{\gamma}\\
	\end{split}
\end{align}
where the physical state of the kinematic model considered in this study using $x(t) =(\chi(t),\gamma(t))^T$, and the input command $u(t)=(\phi(t),n_z(t))^T$ signifies the vector containing roll angle $\phi(t)$ and normal overload $n_z(t)$ along the z-axis direction.
$d = (d_{\chi}, d_{\gamma})^T$ is the disturbance term caused by factors such as wind field and model simplification applied to $\dot{\chi}(t)$ and $\dot{\gamma}(t)$ respectively.  Here the perturbation $d$ is given in the form of sinusoidal noise signal as
\begin{equation}
	\label{equ:d}
	d_{\chi} = L_{d_{\chi}}\sin(\omega_{\chi}t), d_{\gamma} = L_{d_{\gamma}}\cos(\omega_{\gamma}t)
\end{equation}
\begin{table}[hbtp]
	\caption{Hyperparameter declarations}
	\label{tab:hyperparameters}
	\centering
	\footnotesize
	\begin{tabular}{llll}
		\hline
		\bf{Declaration} & \bf{Param} & \bf{Value} & \bf{Unit}\\ 
		\hline % 横线
		Simulation timespan & $T$ & [0,20] & $\text s$\\
		Acceleration of gravity & $g$ & 9.81 & $\text m/\text s^2$ \\
		Consolidated velocity & $V$ & 25 & $\text m/ \text s$ \\
		Initial climb angle & $\gamma(0)$ & $\pi$/4& $\text{rad} $\\
		Initial azimuth angle & $\chi(0)$ & $\pi$/3 & $\text{rad} $\\
		Initial roll angle & $\phi(0)$ & $\pi$/3 & $\text{rad}$\\
		Initial overload & $n_z(0)$ & $1$& $-$\\
		Reference climb angle & $\gamma_c$ & $\pi$/12 & $\text{rad} $\\
		Reference azimuth angle & $\chi_c$ & 0& $\text{rad}$\\
		Reference roll angle & $\phi_c$ & 0& $\text{rad}$\\
		Reference overload & $n_{zc}$ & 0& $-$\\
		Lipschitz constant of $d_{\chi}$ & $L_{d_{\chi}}$ & 0.1 & $-$ \\
		Lipschitz constant of $d_{\gamma}$ & $L_{d_{\gamma}}$ & 0.1 & $-$ \\
		Disturbance frequency of $d_{\chi}$ & $\omega_{\chi}$ & 0.15 & $\text{rad}/ \text s$\\
		Disturbance frequency of $d_{\gamma}$ & $\omega_{\gamma}$ & 0.15 & $\text{rad}/ \text s$\\
		The range of $\phi$ & [$\phi_{\min}$,$\phi_{\max}$]& [-$\pi$/4,$\pi$/4] & $\text{rad} $\\
		The range of $\dot{\phi}$ & [$\dot{\phi}_{\min}$,$\dot{\phi}_{\max}$]& [-$\pi$/6,$\pi$/6] & $\text{rad} $/$\text{s}$\\
		The range of $n_z$ & [$n_{z,\min}$,$n_{z,\max}$]& [-2.1,2.1] & $-$\\
		The range of $\dot{n}_z$ & [$\dot n_{z,\min}$,$\dot n_{z,\max}$]& [-1,1] & $/s$\\
		\hline 
	\end{tabular}
\end{table}
We declare critical hyperparameters for this experiment in TABLE \ref{tab:hyperparameters}. It is stipulated that the initial derivative of all states are zero.
Subsequently, we respectively utilize the underlying:
\begin{itemize}
	\item Average Integral Time Absolute Error(ITAE, the average of the absolute differences between the actual signal and the reference signal, integrated over a specific time period);
	\item Peak Time(PT, the time it takes for a signal to rise from a defined initial value to its highest point);
	\item Maximum Overshoot (MO, the maximum deviation by which a response exceeds its final value) 
	\item Mean value after Stabilization (MS, the mean value of convergence errors in the last 1/4 time interval)
	\item Standard deviation after Stabilization (ST, the standard deviation of convergence errors in the last 1/4 time interval)
\end{itemize}
to quantify the performances of MIMO-PI controller under different circumstances. Assuming that the velocity of aircraft is well-maintained to $V$, and $g$ denotes the gravitational acceleration, which equals to $\mathrm{9.81m/s^2}$. 
Define the reference tracking signals as $x_c = (\chi_c,\gamma_c)^T$ with the tracking error $e(t) = (e_{\chi}(t),e_{\gamma}(t))^T$ as
\begin{align}
	\begin{split}
		e(t) = x_c - x(t)= (\chi_c - \chi(t), \gamma_c - \gamma(t))^T
	\end{split}
\end{align}
Under this specific form of error, the above perturbed system under the constrain $\dot{x}_c = 0$ is transformed into:
\begin{align}
	\label{eq:aircraft_error_model}
	\dot{e}(t) = f_e(e(t),u(t)) + d_e 
\end{align}
where $d_e = (-d_{\chi}, - d_{\gamma} )^T$ and
\begin{align}
	f_e(e(t),u(t)) = 
	\begin{pmatrix}
		-\frac{g\tan\phi(t)}{V}\\
		-\frac{g(n_z(t)\cos \phi(t) - \cos(\gamma_c - e_{\gamma}(t)))}{V}
	\end{pmatrix}
\end{align}
The Jacobians of $f_e(e,u)$  with respect to $e$ and $u$ are:
\begin{align}
	&\frac{\partial f_e(e,u)}{\partial e}
	=
	\begin{pmatrix}
		0 & 0\\
		0 & \frac{g}{V}\sin(\gamma_c - e_{\gamma})
	\end{pmatrix} \\
	 &\frac{\partial f_e(e,u)}{\partial u} =\begin{pmatrix}
		-\frac{g}{V}\sec^2\phi &  0\\
		\frac{gn_z}{V}\sin\phi & -\frac{g}{V}\cos \phi
	\end{pmatrix}
\end{align}
The Jacobians of the above equation at the equilibrium point \(e_{\gamma} = 0\), \(e_{\chi} = 0\), \(\phi = 0\), and \(n_z = \cos\gamma_c/g \) are:
\begin{align}
	\frac{\partial f_e(0)}{\partial e}
	=
	\begin{pmatrix}
		0 & 0\\
		0 & \frac{g}{V}\sin\gamma_c
	\end{pmatrix}
	,\ \frac{\partial f_e(0)}{\partial u} =\begin{pmatrix}
		-\frac{g}{V} &  0\\
		0 & -\frac{g}{V}
	\end{pmatrix}
\end{align}
For stabilizing the above perturbed system, we design a MIMO-PI controller in the following form as
\begin{align}
	u(t) = K_Pe(t) + K_I\int_0^t e(t) dt
\end{align}
The optimal $K_P^*$ and $K_I^*$ computed by solving the Eq.(\ref{eq:optimal_model}) using GA optimization method are:
\begin{align}
	\label{eq:K2}
	\begin{split}
		K_P^* = \begin{pmatrix}
			1.6968  & 0.5906 \\
			-0.5906  &  1.9556
		\end{pmatrix}\\
		K_I^* = \begin{pmatrix}
			3.4869  & 0.1784 \\
			-0.1784  &  3.4869
		\end{pmatrix}
	\end{split}
\end{align}
where the corresponding convergence indictors are as follows
\begin{align}
	R_K^* = 0.4427,\ I_K^* = 4.9244
\end{align}
\subsubsection{The effect of error stabilization}
\begin{figure}[htbp]
	\centering
	\begin{tabular}{cc}
		\includegraphics[width=0.22\textwidth]{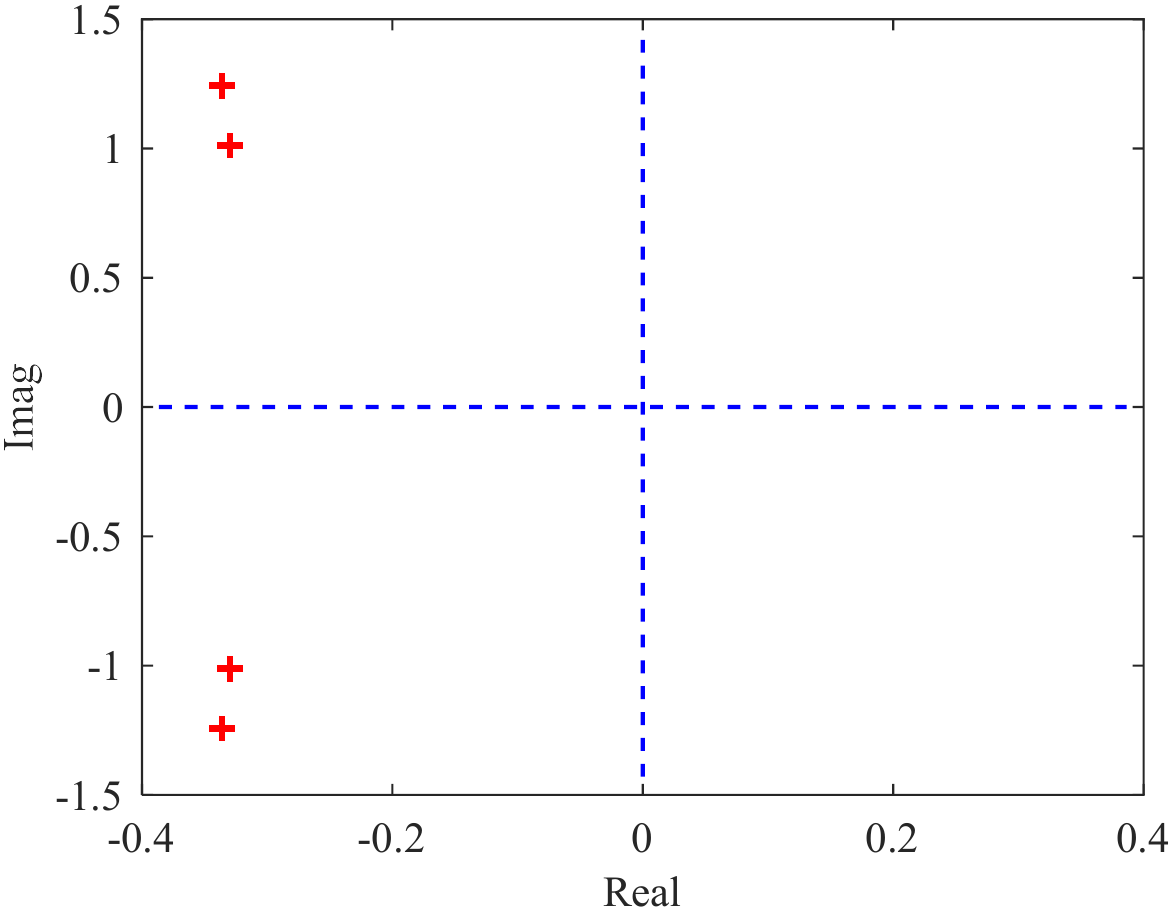} &
		\includegraphics[width=0.22\textwidth]{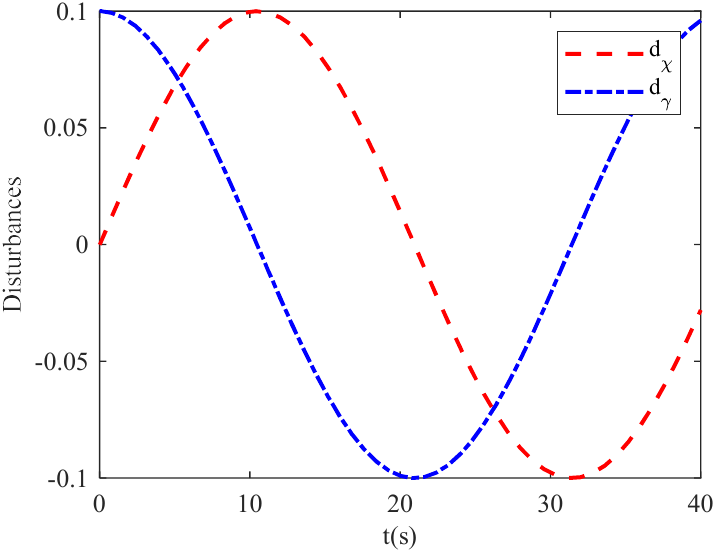} \\
		(a) & (b)
	\end{tabular}
	\caption{The eigenvalue distribution at the origin and the applied sinusoidal disturbance $d_{\chi}$, $d_{\gamma}$. (a): Eigenvalue distribution; (b): $d_{\chi}$, $d_{\gamma}$.}
	\label{Fig:6}
\end{figure}

\begin{figure}[!t]
	\centering
	\begin{tabular}{cccc}
		\includegraphics[width=0.23\textwidth]{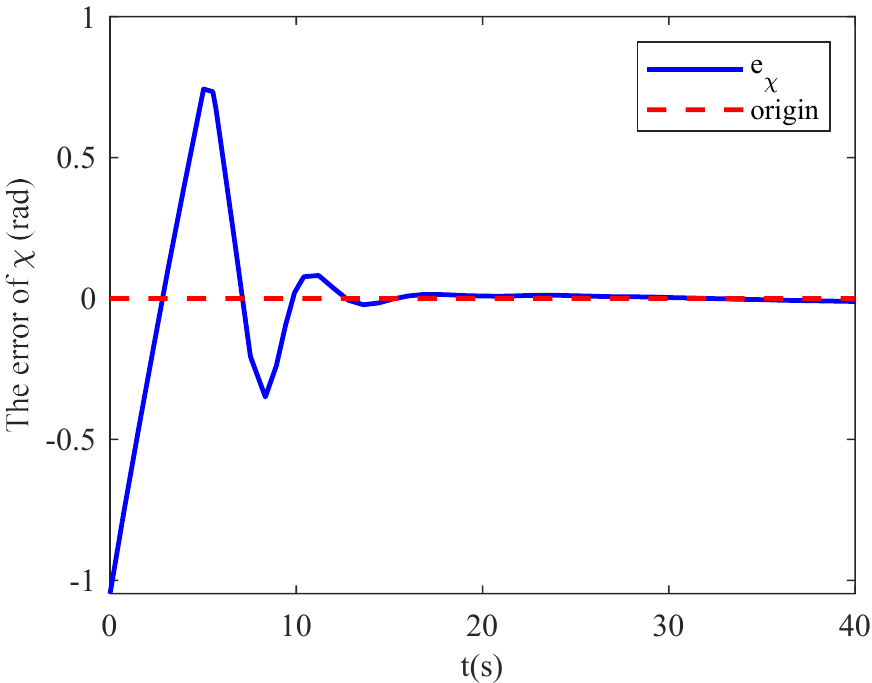} &
		\includegraphics[width=0.23\textwidth]{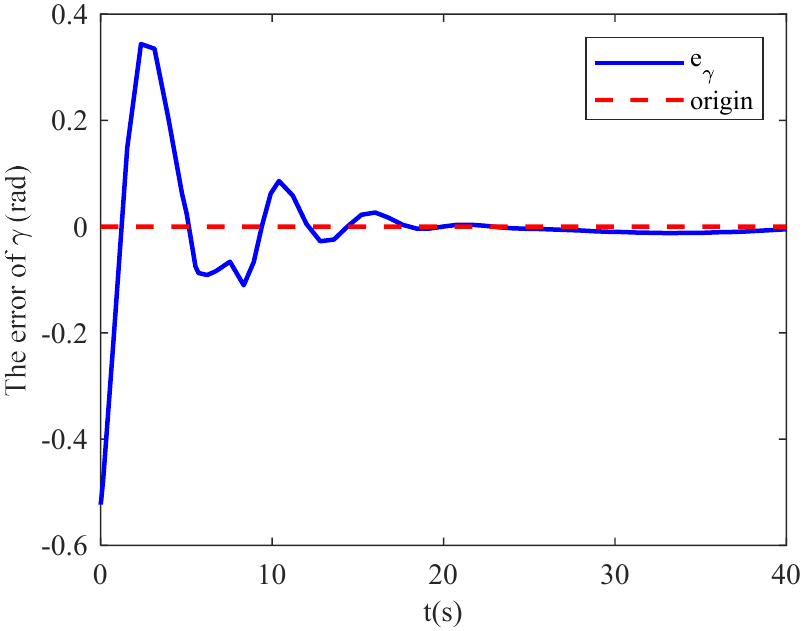} \\
		(a) & (b) \\
		\includegraphics[width=0.23\textwidth]{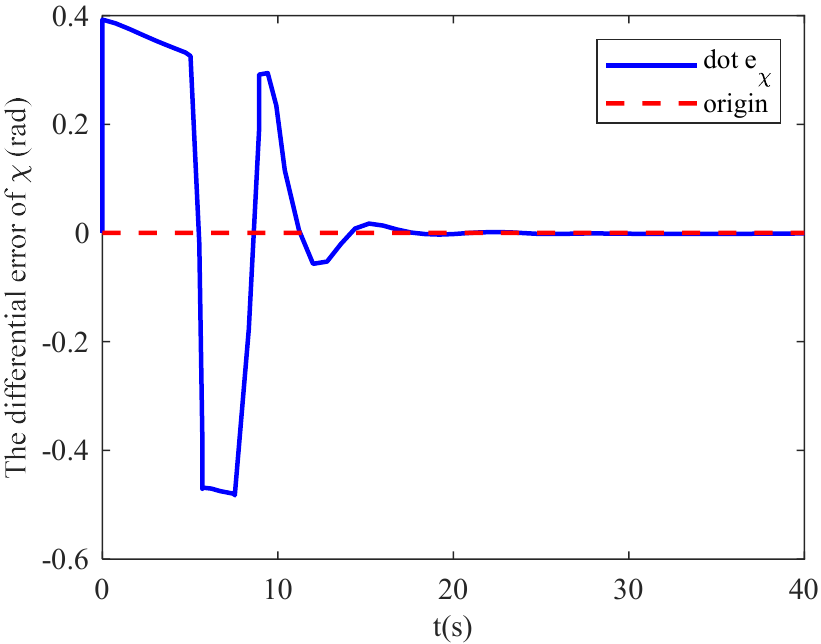} &
		\includegraphics[width=0.23\textwidth]{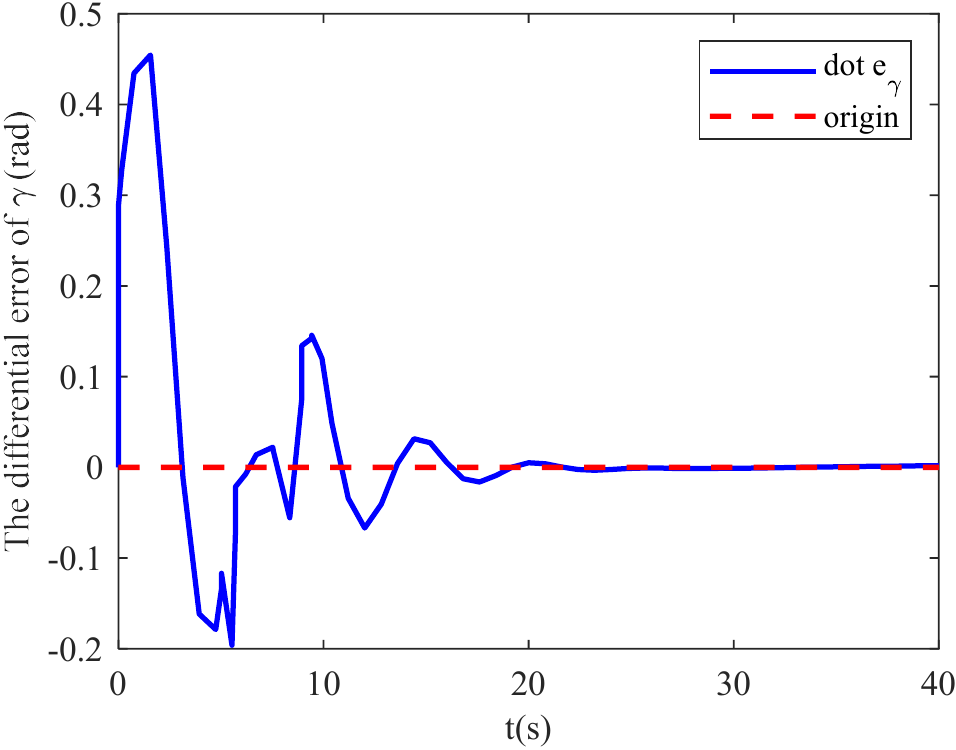}\\
		(c) & (d)\\
	\end{tabular}
	\caption{Error stabilization profiles comparsion under the MIMO-PI controller using $R_K^*$ and $I_K^*$ for $e_{\gamma}$, $e_{\chi}$, $\dot{e}_{\gamma}$ and $\dot{e}_{\chi}$. 
		(a): $e_{\chi}(t)$ profile; 
		(b): $e_{\gamma}(t)$ profile;
		(c): $\dot{e}_{\chi}(t)$ profile;
		(d): $\dot{e}_{\gamma}(t)$ profile.
	}
	\label{Fig:5}
\end{figure}

\begin{figure}[!t]
	\centering
	\begin{tabular}{cc}
		\includegraphics[width=0.23\textwidth]{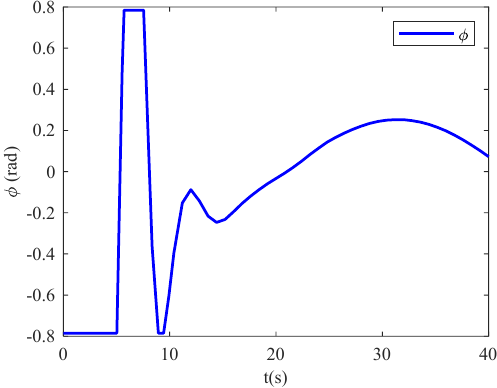} &
		\includegraphics[width=0.23\textwidth]{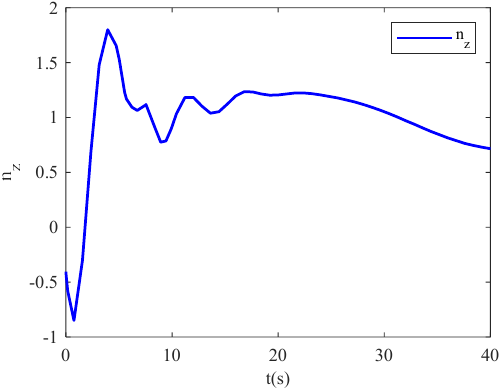} \\
		(a) & (b) \\
	\end{tabular}
	\caption{Input commands for MIMO-PI controller within coefficients $R_K^*$ and $I_K^*$.
		(a): $\phi(t)$ profile; 
		(b): $n_z(t)$ profiles.}
	\label{Fig:12}
\end{figure}

In the case of the current optimal coefficients \( K_p^* \) and \( K_I^* \), the eigenvalue distribution of \( A_K(0) \) corresponding to the MIMO-PI controller is shown in Figure \ref{Fig:6}-(a), from which it can be observed that all eigenvalues lie in the left half of the complex plane. The applied sinusoidal disturbance signal is presented in Figure \ref{Fig:6}-(b).  
Under the MIMO-PI controller corresponding to \( K_P^* \) and \( K_I^* \), the time-series curves of tracking errors \( e_{\chi} \), \( e_{\gamma} \), \( \dot{e}_{\chi} \), and \( \dot{e}_{\gamma} \) are illustrated in Figure \ref{Fig:5}. It can be observed that \( e_{\chi} \) and \( e_{\gamma} \) rapidly converge to a very small region near the origin within 20 seconds. During this process, the maximum overshoot of \( e_{\chi} \) does not exceed 0.8 rad, and that of \( e_{\gamma} \) is within 0.4 rad, with their maximum peak time less than 4 seconds. It is also noteworthy that the input command curves under these controller parameters are presented in Figure \ref{Fig:12}, which indicate that the commands satisfy the corresponding input constraints.  

\subsubsection{The influence of $R_K$ and $I_K$}
% It’s worth pointing out that the results in Fig.\ref{Fig:7} indicates that compensation method can achieve a competitive average reduction of approximately 28$\%$ in ITAE and 43$\%$ in MO. Notably, these performance enhancements are frequently attributed to the utilization of first-order differentials, which effectively minimize errors convergence rates. Although this may result in a slight increase in the PT indicator, the magnitude of this increase is typically insignificant.
To validate the rationality of using indices \( R_K \) and \( I_K \) for quantifying robust stability during error convergence, we derive new controller coefficients \( K \) by introducing incremental perturbations \( \Delta K \) to the baseline optimal coefficients \( K^* = (K_P^*, K_I^*) \). 
\begin{align}
	K = K^* + \Delta K 
\end{align}
We then compare the corresponding response curves and their associated performance metrics—ITAE, MO, PT, MS, and ST—across multiple such coefficient sets. For simplicity, $\Delta K$ can be described in the following form
\begin{equation}
	\label{equ:DK}
	\Delta K = -\varepsilon (I_p,I_i)
\end{equation}
where $I_p$ and $I_i$ both are identity matrixes for standard form $m=n$, and $\epsilon \in \mathbb{R}^1$ is served as a regulation variable to determine $\Delta K$.
\begin{table}[htbp]
	\caption{The different $\varepsilon$ and corresponding $R_K$, $I_K$.}
	\label{tab:varepsilon}
	%\centering
	\begin{tabular}{llll|llll}
		\hline
		\bf{Type} & \bf{$\varepsilon$} & \bf{$R(f)$} & \bf{$I(f)$}
		 &
		\bf{Type} & \bf{$\varepsilon$} & \bf{$R(f)$} & \bf{$I(f)$}\\ 
		\hline % 横线
		$K_1$ & -4 & 0.7742 & 7.873 
		& 
		$K_4$ & 0.5 & 0.3626 & 5.207\\
		$K_2$ & -2 & 0.6754 & 6.057
		&
		$K_5$ & 0.8 & 0.2849 & 6.161\\
		$K_3$ & -1 & 0.5913 & 5.312
		&
		$K_6$ & 1 & 0.2192 & 8.258\\    
		\hline 
	\end{tabular}
\end{table}

\begin{figure*}[htbp]
	\centering
	\begin{tabular}{cc}
		\includegraphics[width=0.46\textwidth]{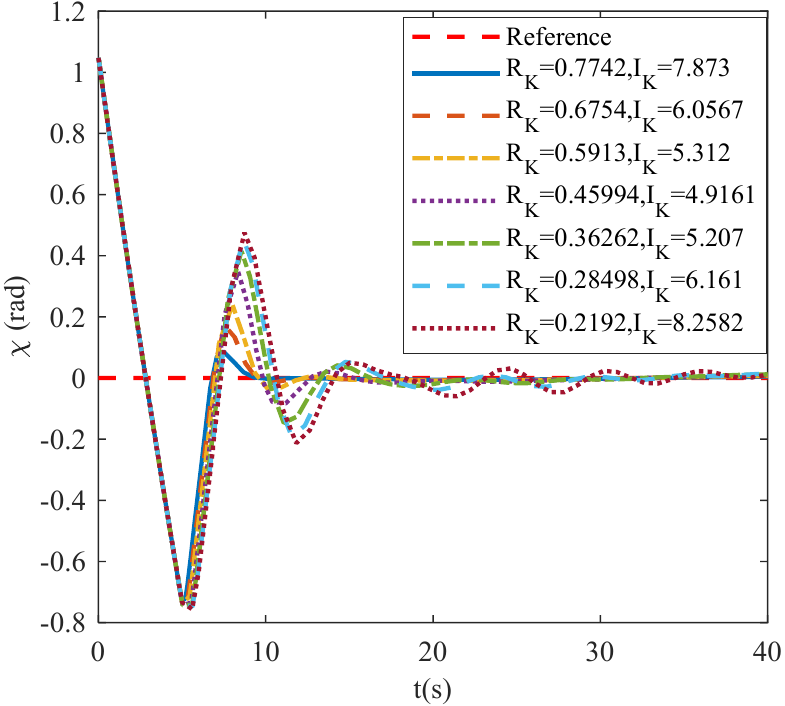} &
		\includegraphics[width=0.46\textwidth,height=0.85\columnwidth]{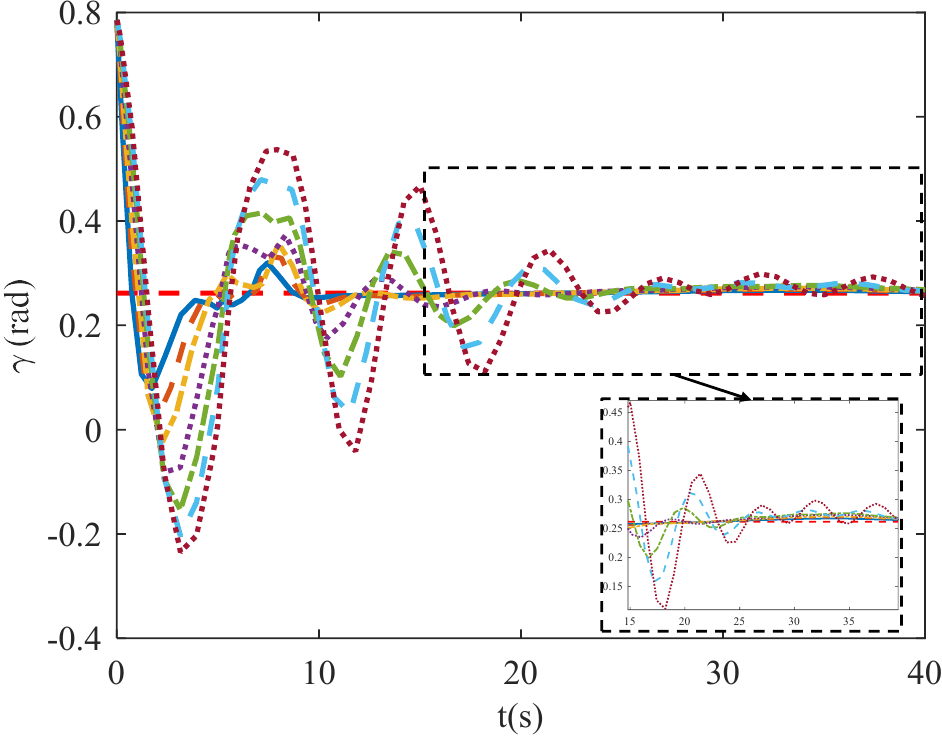} \\
		(a) & (b) \\
		\includegraphics[width=0.46\textwidth]{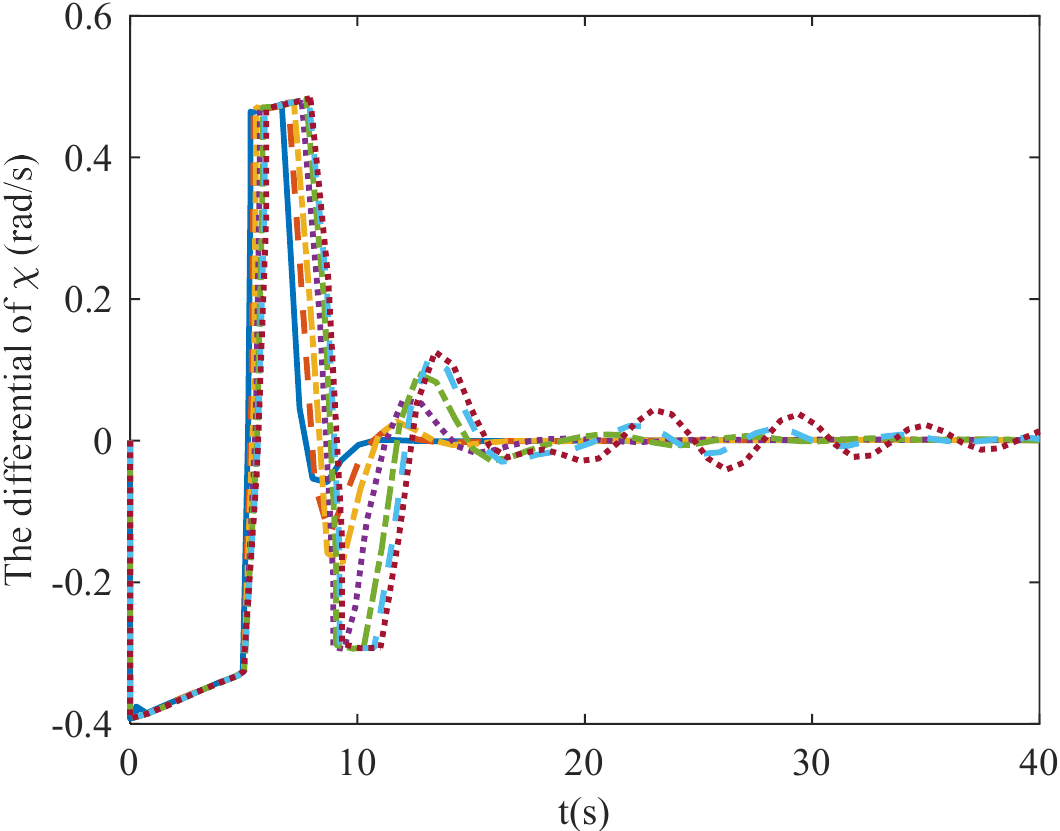} &
		\includegraphics[width=0.46\textwidth]{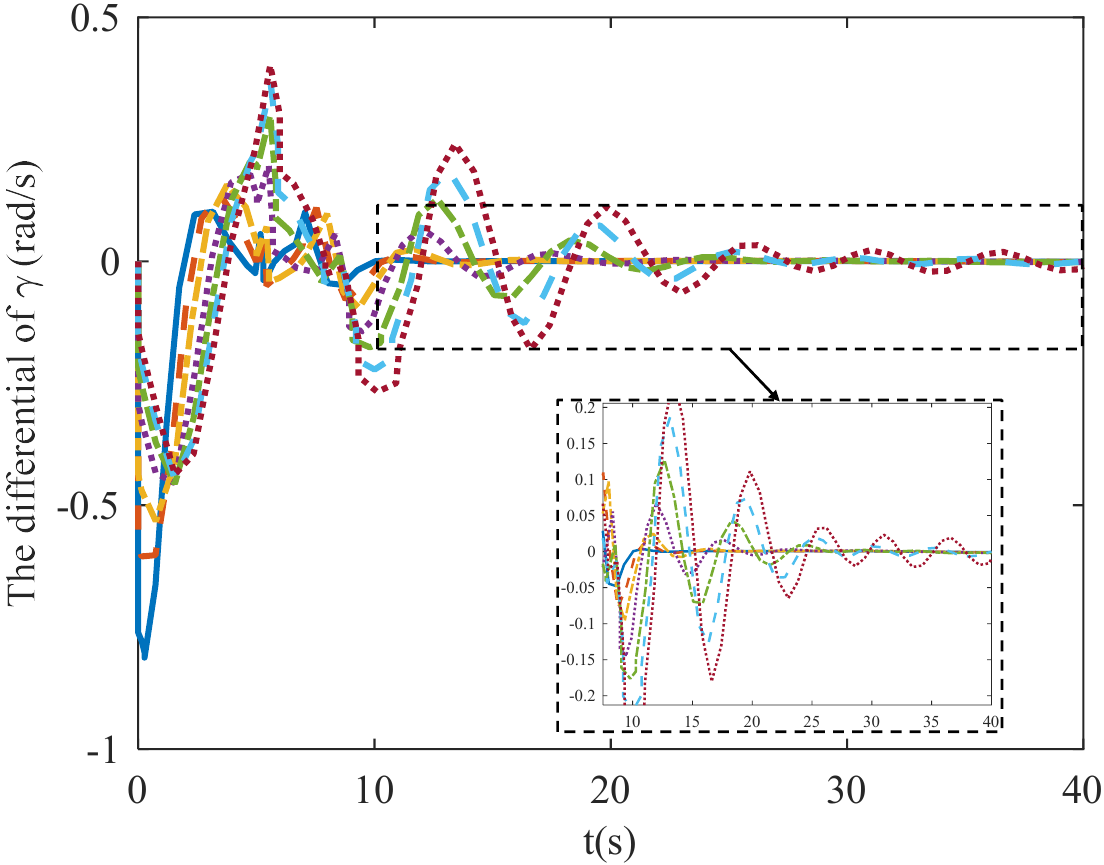} \\
		(c)  & (d) 
	\end{tabular}
	\caption{Profiles comparsion of $\gamma$, $\chi$, $\dot{\gamma}$ and $\dot{\chi}$ accompanied by different $\varepsilon$ within coresponding $R_K$ and $I_K$.
		(a): Comparison of response profiles for $R_K, I_K$ on $\chi$; 
		(b): Comparison of response profiles for $R_K, I_K$ on $\gamma$;
		(c): Comparison of response profiles for $R_K, I_K$ on $\dot{\chi}$;
		(d): Comparison of response profiles for $R_K, I_K$ on $\dot{\gamma}$.
	}
	\label{Fig:8}
\end{figure*}

Subsequently, we analyzed the dynamic response, as depicted in Fig.\ref{Fig:8}. The result indicates that across multiple channels, a higher $R_K$ correlates with superior exponential convergence. Conversely, it’s more likely to cause oscillation and a lack of convergence when a small $R_K$ occurs. This is because a larger \( R_K \) tends to indicate a higher exponential convergence rate, causing the error energy to decay more rapidly throughout the convergence process—specifically, with smaller amplitudes and slower oscillation frequencies. In a sense, \( R_K \) exerts a significantly more pronounced influence on steady-state error and convergence performance than \( I_K \).

\begin{figure*}[htbp]
	\centering
	\begin{tabular}{ccc}
		\includegraphics[width=0.32\textwidth]{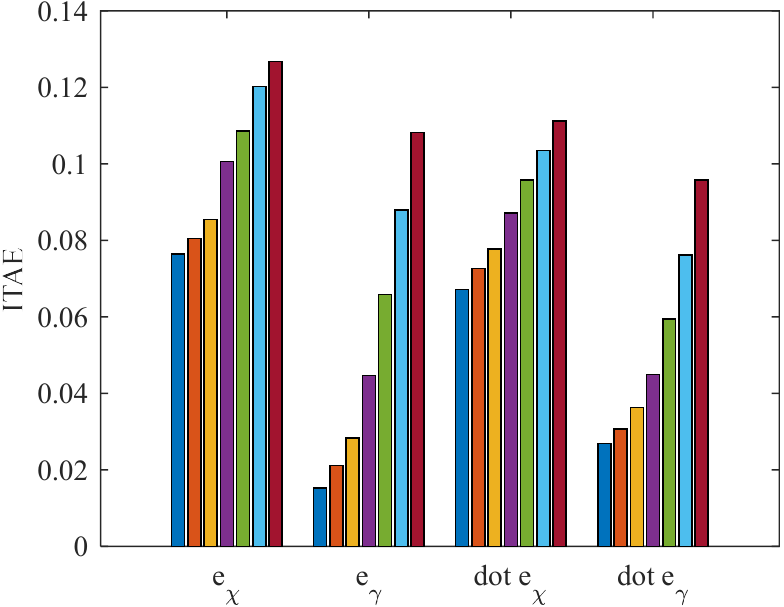} &
		\includegraphics[width=0.32\textwidth]{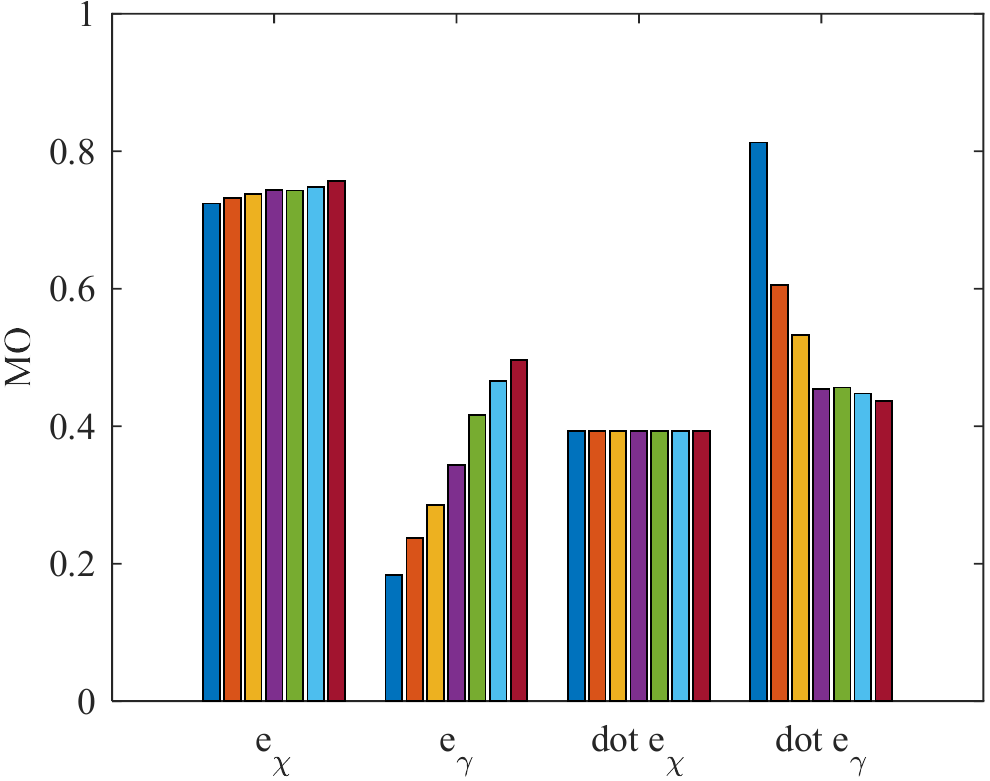} &
		\includegraphics[width=0.32\textwidth]{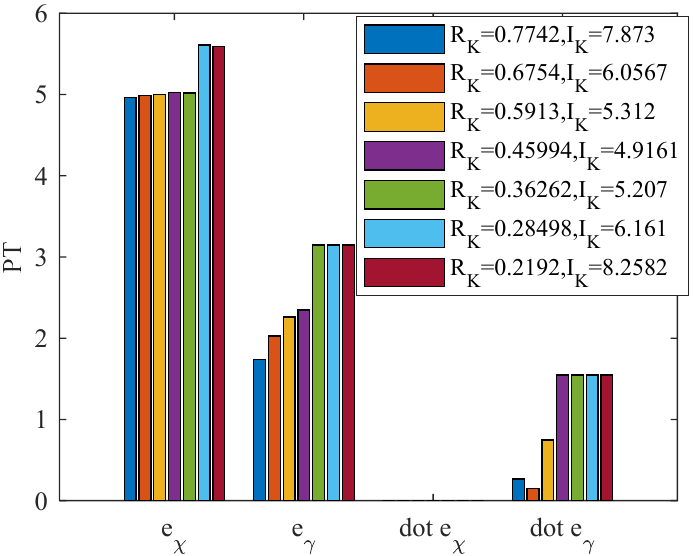}\\
		(a)  & (b) & (c)  \\
	\end{tabular}
	\caption{Performance indicators comparison of ITAE, MO and PT between multiple corresponding $R_K, I_K$ respectively for $e_{\gamma}$, $e_{\chi}$, $\dot{e}_{\gamma}$ and $\dot{e}_{\chi}$. 	
		(a): The comparison of ITAE; 
		(b): The comparison of MO;
		(c): The comparison of PT.	}
	\label{Fig:7}
\end{figure*}

\begin{figure*}[htbp]
	\centering
	\begin{tabular}{cc}
		\includegraphics[width=0.48\textwidth]{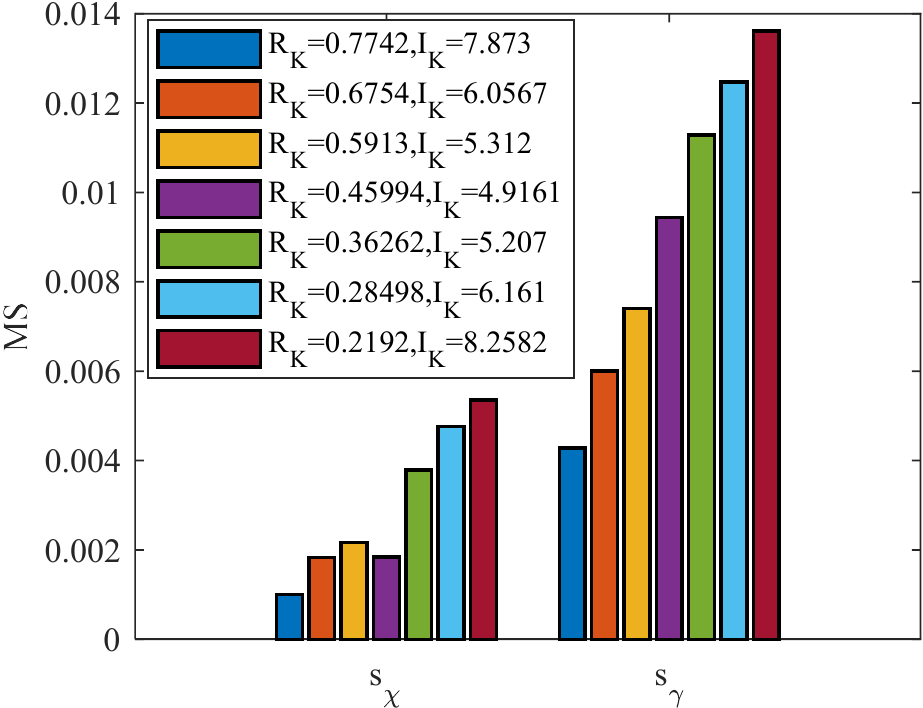} &
		\includegraphics[width=0.48\textwidth]{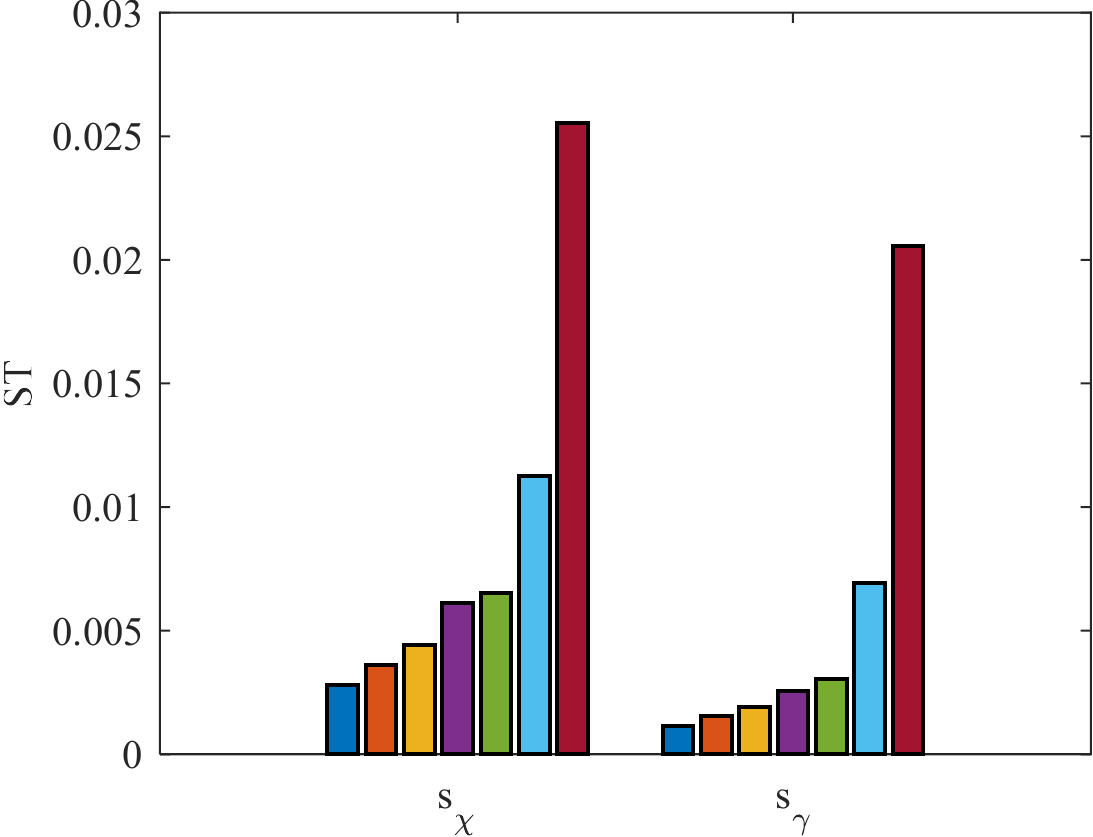}\\
		(a) & (b) 
	\end{tabular}
	\caption{Comparison of MS and ST for states \( s_{\chi} \) and \( s_{\gamma} \) under robust indices \( R_K \) and \( I_K \) corresponding to different \( \varepsilon \). (a): Comparison of MS; (b): Comparison of ST.}
	\label{Fig:13}
\end{figure*}

\begin{figure*}[htbp]
	\centering
	\begin{tabular}{cccc}
		\includegraphics[width=0.23\textwidth]{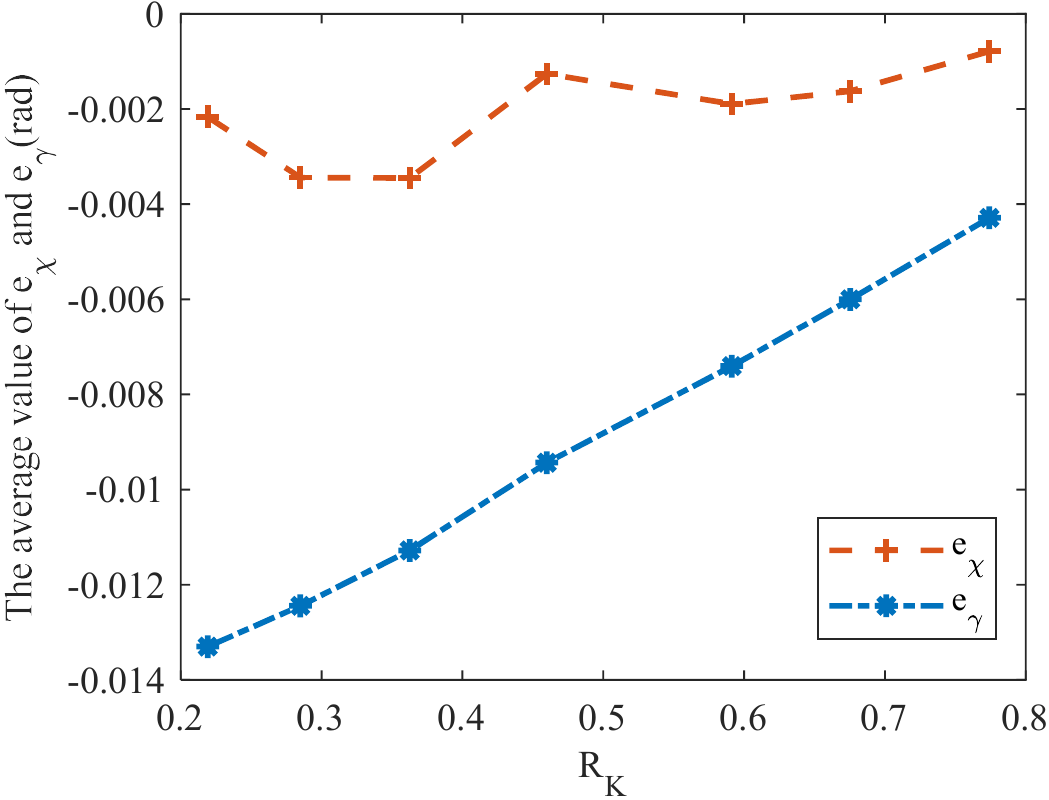} &
		\includegraphics[width=0.23\textwidth]{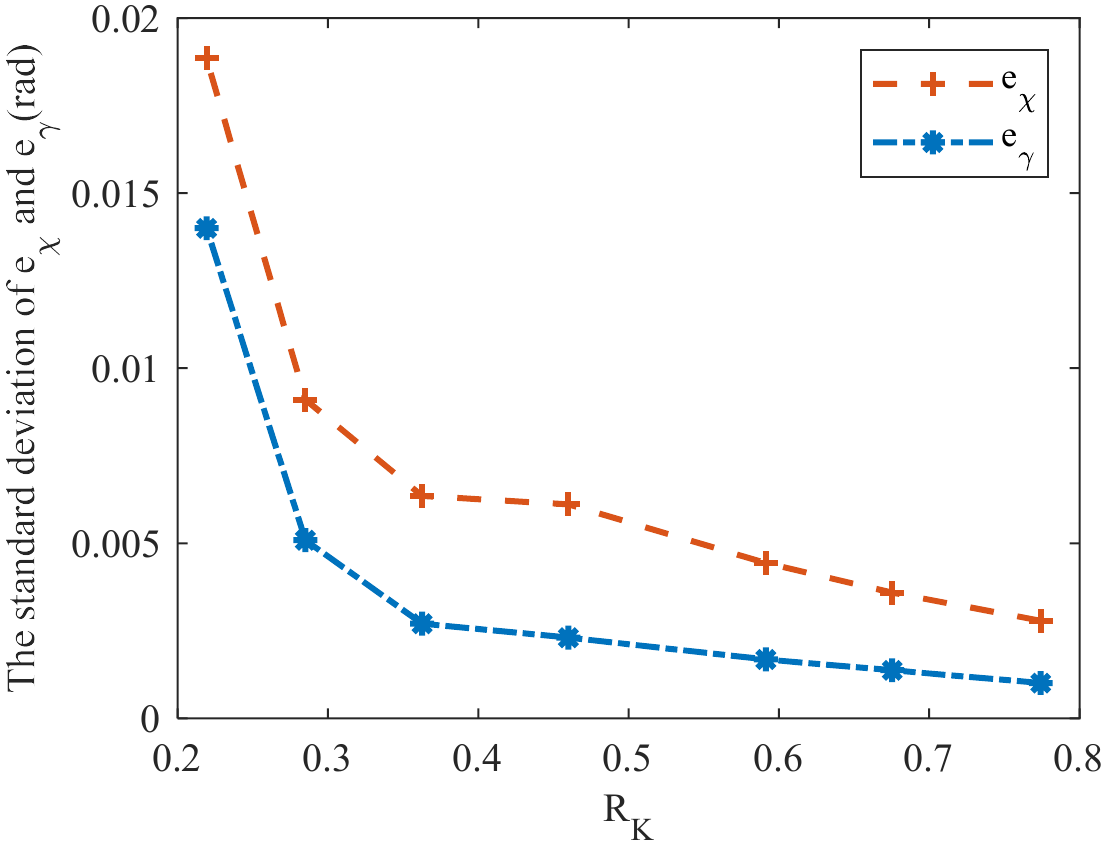} &
		\includegraphics[width=0.23\textwidth]{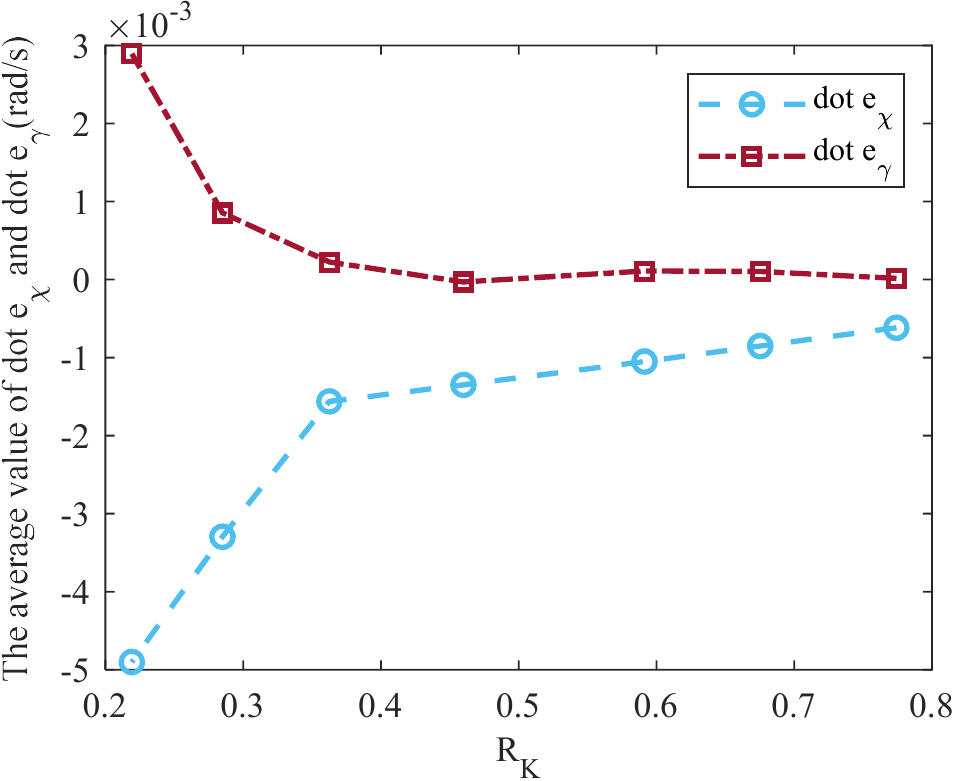} &
		\includegraphics[width=0.23\textwidth]{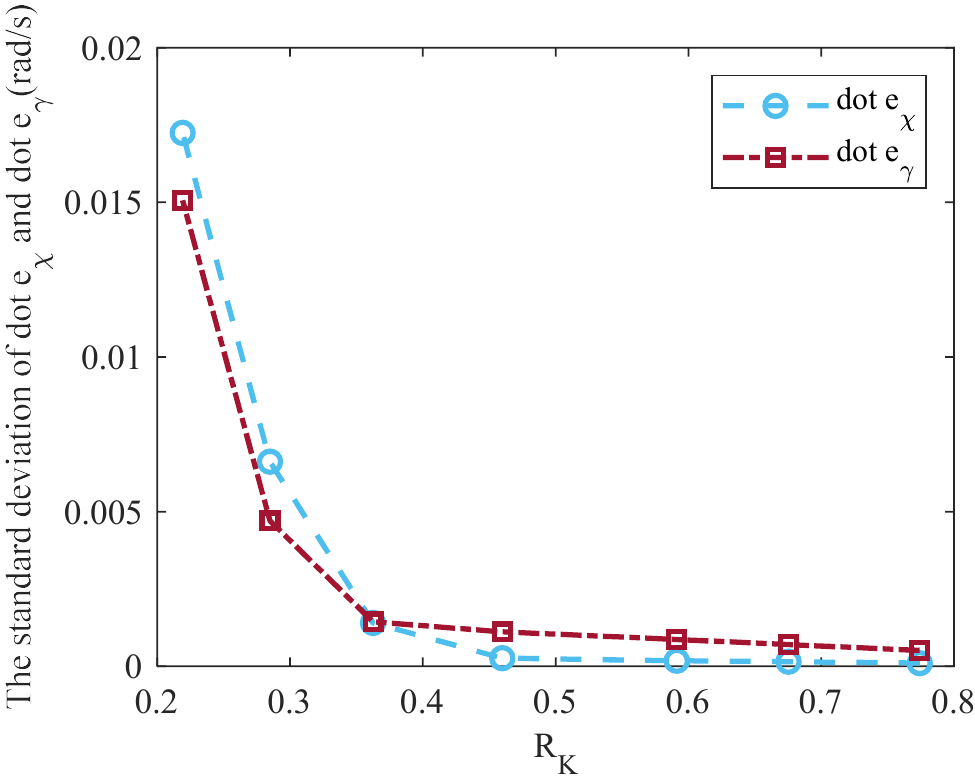}\\
		(a) & (b) & (c) & (d) \\
		\includegraphics[width=0.23\textwidth]{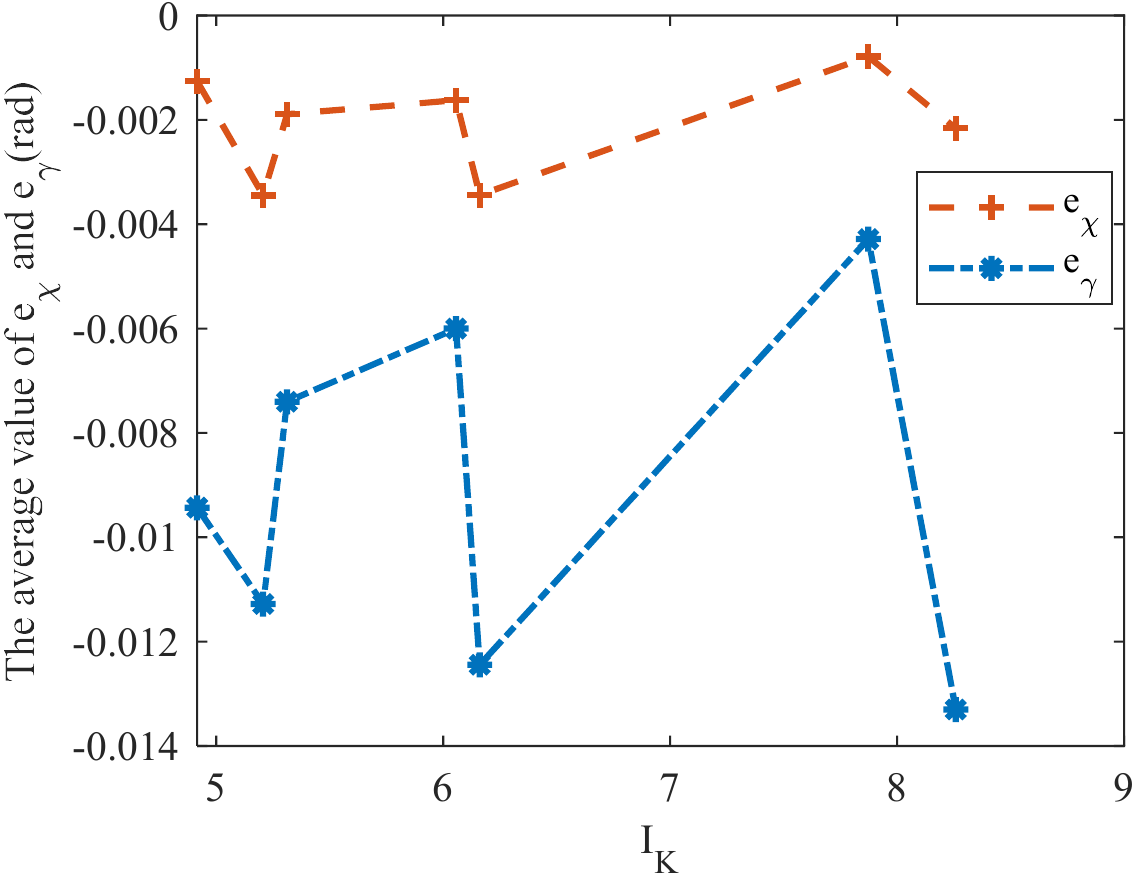} &
		\includegraphics[width=0.23\textwidth]{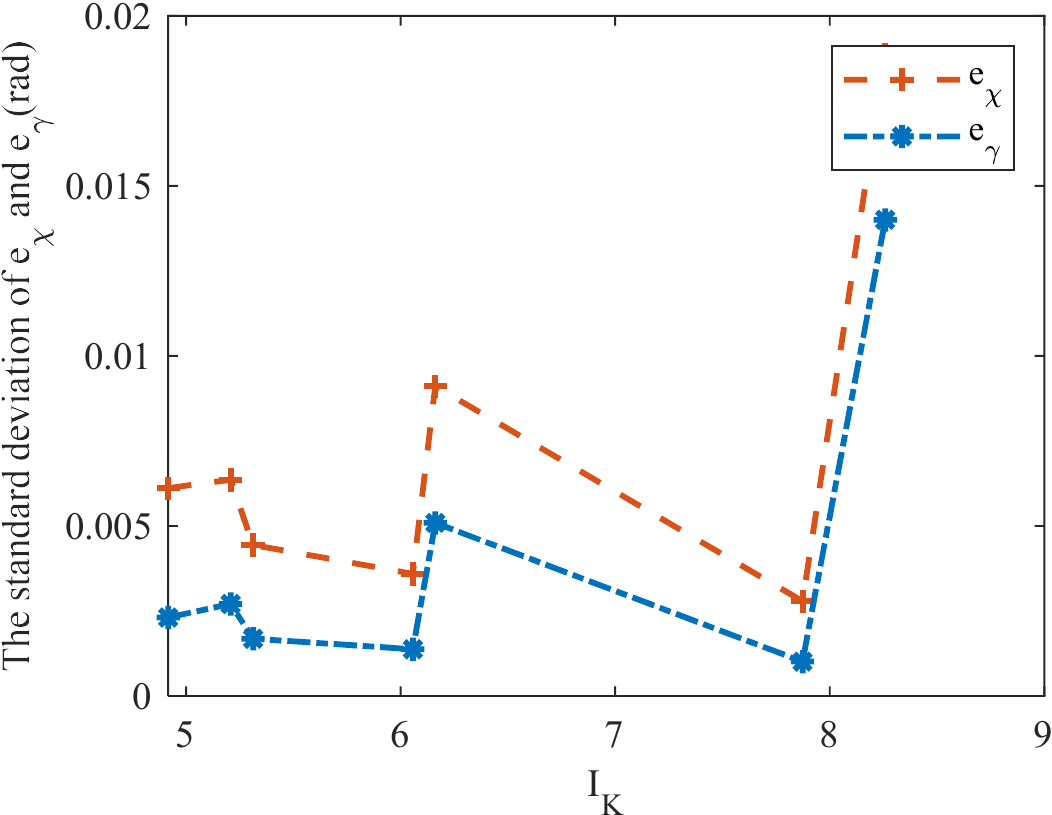} &
		\includegraphics[width=0.23\textwidth]{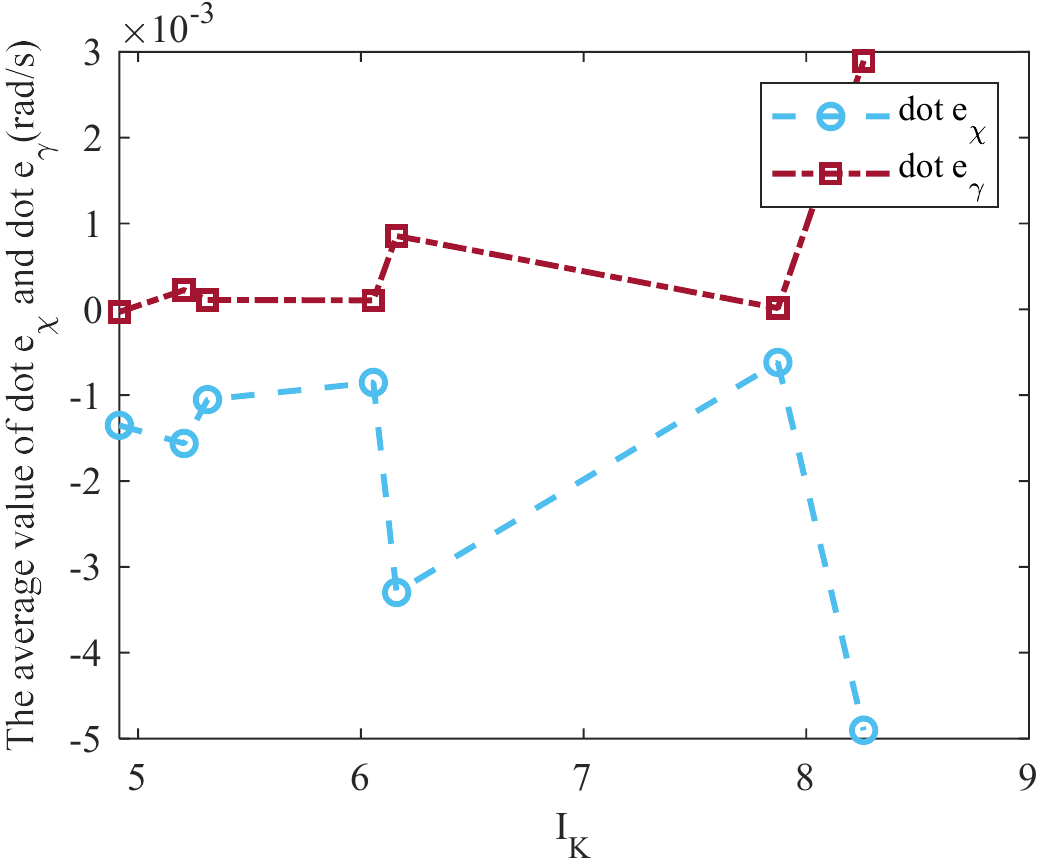} &
		\includegraphics[width=0.23\textwidth]{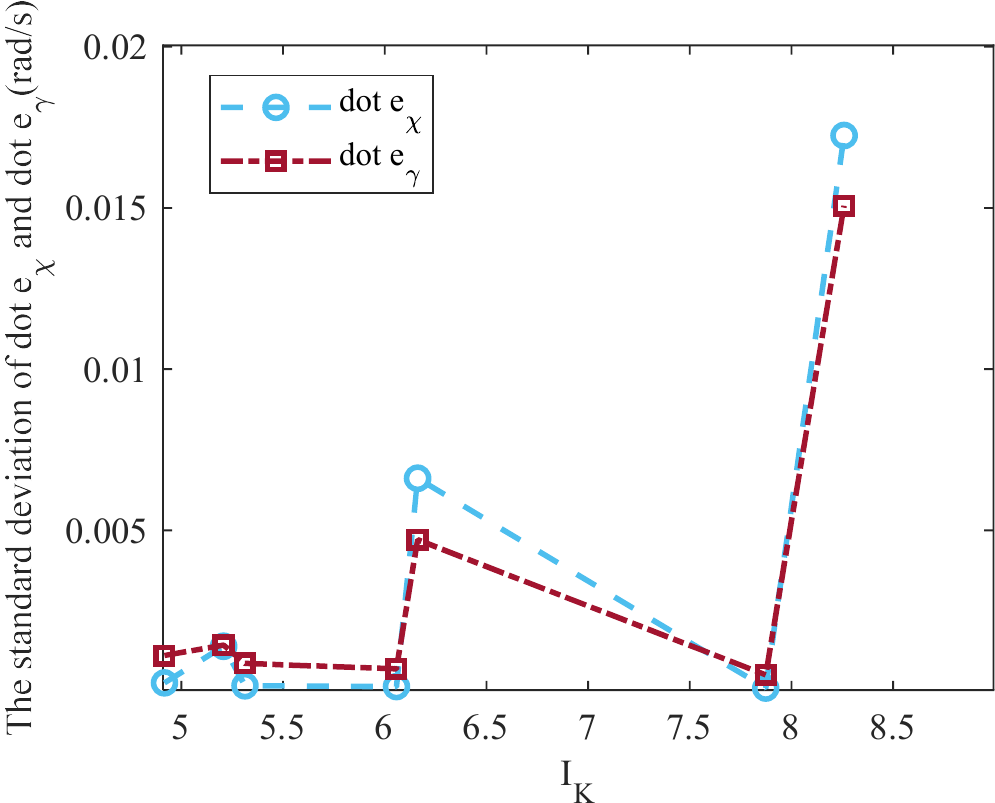}\\
		(e) & (f) & (g) & (h) 
	\end{tabular}
	\caption{This figure illustrates the relationships of the average value and standard deviation indices for \( e_{\gamma} \), \( e_{\chi} \), \( \dot{e}_{\gamma} \), and \( \dot{e}_{\chi} \) with respect to the variations of \( R_K \) and \( I_K \). 
	(a): The average values comparsion of \( e_{\gamma} \) and \( e_{\chi} \) with respect to \( R_K \);
	(b): The standard deviations comparsion of \( e_{\gamma} \) and \( e_{\chi} \) with respect to \( R_K \);
	(c): The average values comparsion of \( \dot e_{\gamma} \) and \(\dot e_{\chi} \) with respect to \( R_K \);
	(d): The standard deviations comparsion of \( \dot e_{\gamma} \) and \( \dot e_{\chi} \) with respect to \( R_K \);
	(e): The average values comparsion of \( e_{\gamma} \) and \( e_{\chi} \) with respect to \( I_K \);
	(f): The standard deviations comparsion of \( e_{\gamma} \) and \( e_{\chi} \) with respect to \( I_K \);
	(g): The average values comparsion of \( \dot e_{\gamma} \) and \(\dot e_{\chi} \) with respect to \( I_K \);
	(h): The standard deviations comparsion of \( \dot e_{\gamma} \) and \( \dot e_{\chi} \) with respect to \( I_K \).}
	\label{Fig:9}
\end{figure*}

The trend profiles quantitatively illustrating the relationship between the $R_K$ and various performance metrics including ITAE, MO, and PT are depicted in Fig.\ref{Fig:7}, the comparsion of performance MS, ST on 
\begin{align}
	s_{\chi} = \sqrt{e_{\chi}^2 + \dot{e}_{\chi}^2},\ s_{\gamma} = \sqrt{e_{\gamma}^2 + \dot{e}_{\gamma}^2}
\end{align}
are presented in Fig.\ref{Fig:13}. Specifically, higher $R_K$ corresponds to lower ITAE, MS and ST for all errors $e_{\chi}$, $e_{\gamma}$, $\dot e_{\chi}$, $\dot e_{\gamma}$. Furthermore, in most cases, larger \( R_K \) tends to correlate with smaller MO and PT, translating to a reduced maximum overshoot and a shorter peak time. 
Specifically, to explore how \( R_K \) and \( I_K \) influence the convergence processes of \( e_{\chi} \) and \( e_{\gamma} \), we analyzed the effects of \( R_K \) and \( I_K \) respectively on the mean values and standard deviations of the errors \( e_{\chi} \), \( e_{\gamma} \), and their derivatives \( \dot{e}_{\chi} \), \( \dot{e}_{\gamma} \) during convergence, as illustrated in Figure \ref{Fig:9}. 
Notably, comparisons between subfigures (b), (d) and (f), (h) reveal that the influences of \( R_K \) and \( I_K \) on \( e_{\chi} \) and \( e_{\gamma} \) originate from their actions on \( \dot{e}_{\chi} \) and \( \dot{e}_{\gamma} \). As shown by the comparison of subfigures (c) and (g), the effects of \( I_K \) and \( R_K \) on the mean values of the error derivatives \( \dot{e}_{\chi} \) and \( \dot{e}_{\gamma} \) are relatively minor (on the order of \( 10^{-3} \) only). Meanwhile, comparisons of subfigures (d) and (h) indicate that \( R_K \) has a significant impact on the standard deviations of \( \dot{e}_{\chi} \) and \( \dot{e}_{\gamma} \) during convergence—specifically, \( R_K \) exhibits an approximately inverse relationship with these standard deviations. This implies that a larger \( R_K \) can notably reduce oscillations during convergence and enhance the quality of robust convergence.  

In a word, the largest \( R_K \) signifies an overall enhancement of comprehensive performance during the error stabilization process, as a larger \( R_K \) reduces the energy of error derivatives during convergence, mitigates the oscillatory nature of errors in the exponential convergence process, and thus enhances the quality of robust stability.  

\subsubsection{The analysis of robustness  }
\begin{figure*}[htbp]
	\centering
	\begin{tabular}{cc}
		\includegraphics[width=0.46\textwidth]{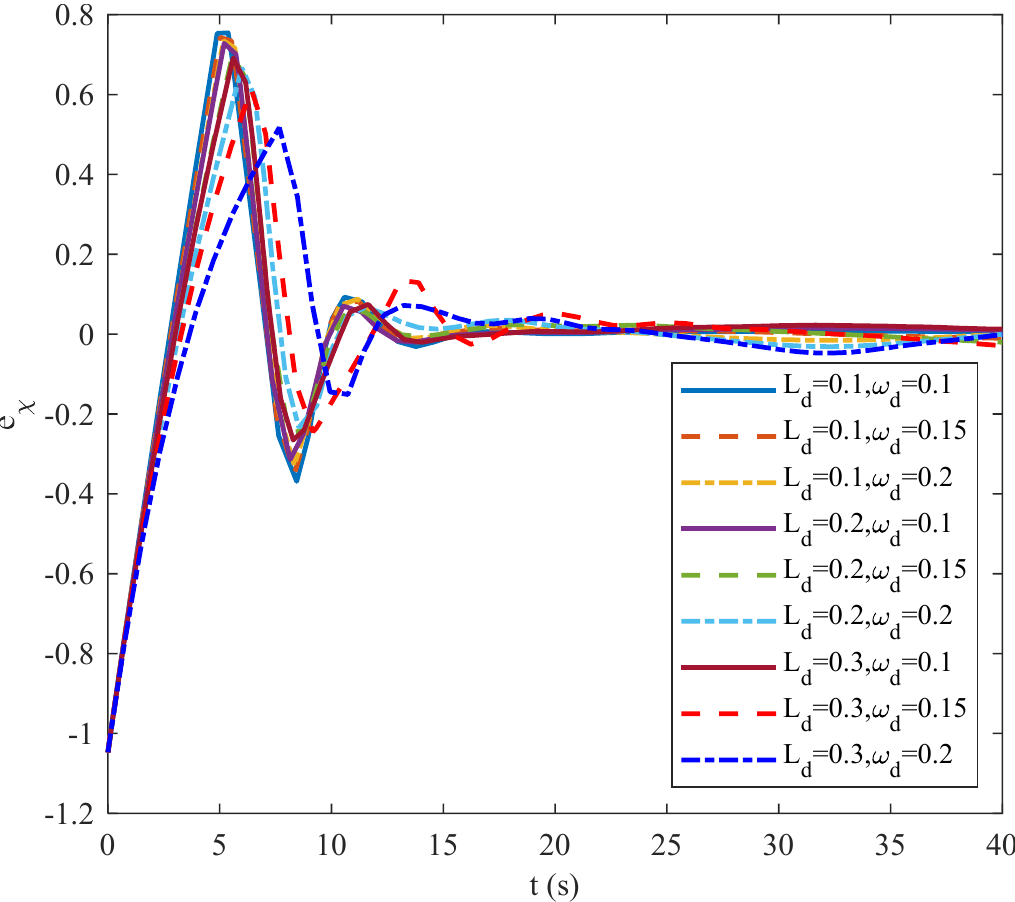} &
		\includegraphics[width=0.46\textwidth,height=0.85\columnwidth]{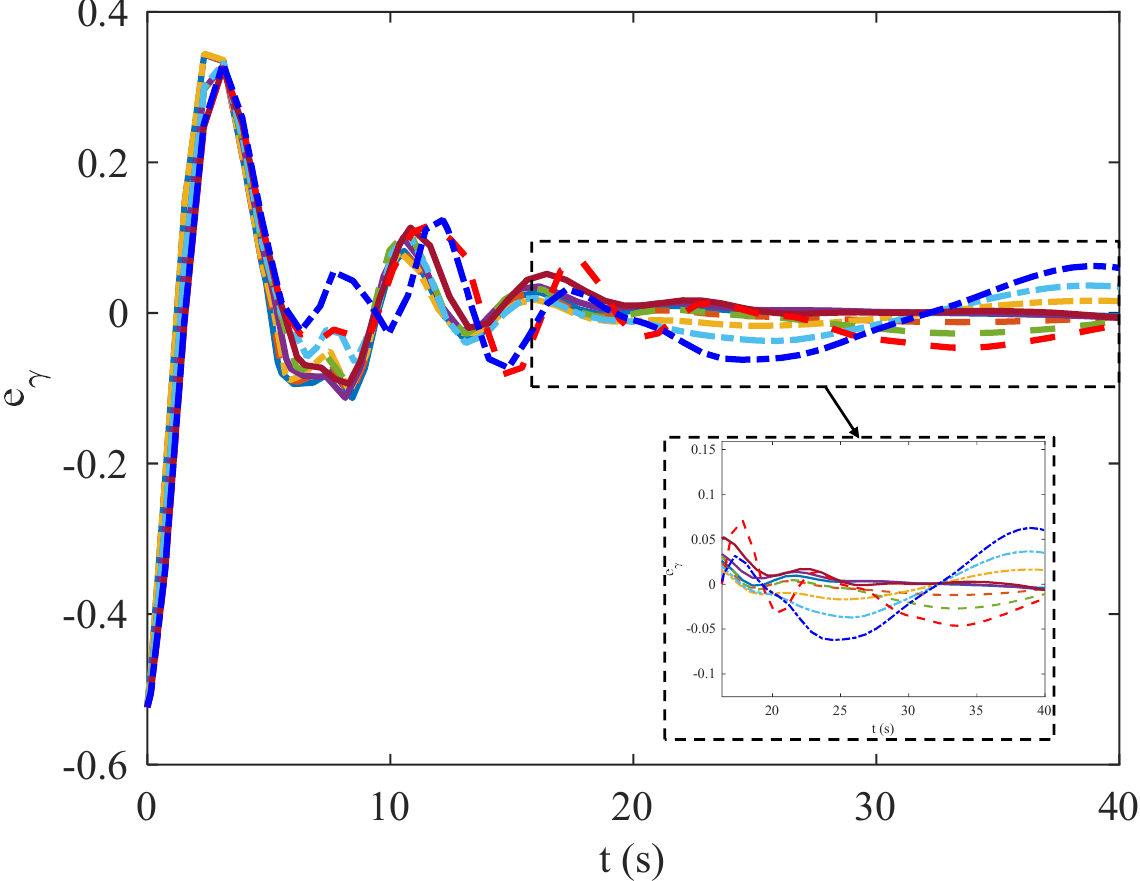} \\
		(a) & (b) \\
		\includegraphics[width=0.46\textwidth,height=0.85\columnwidth]{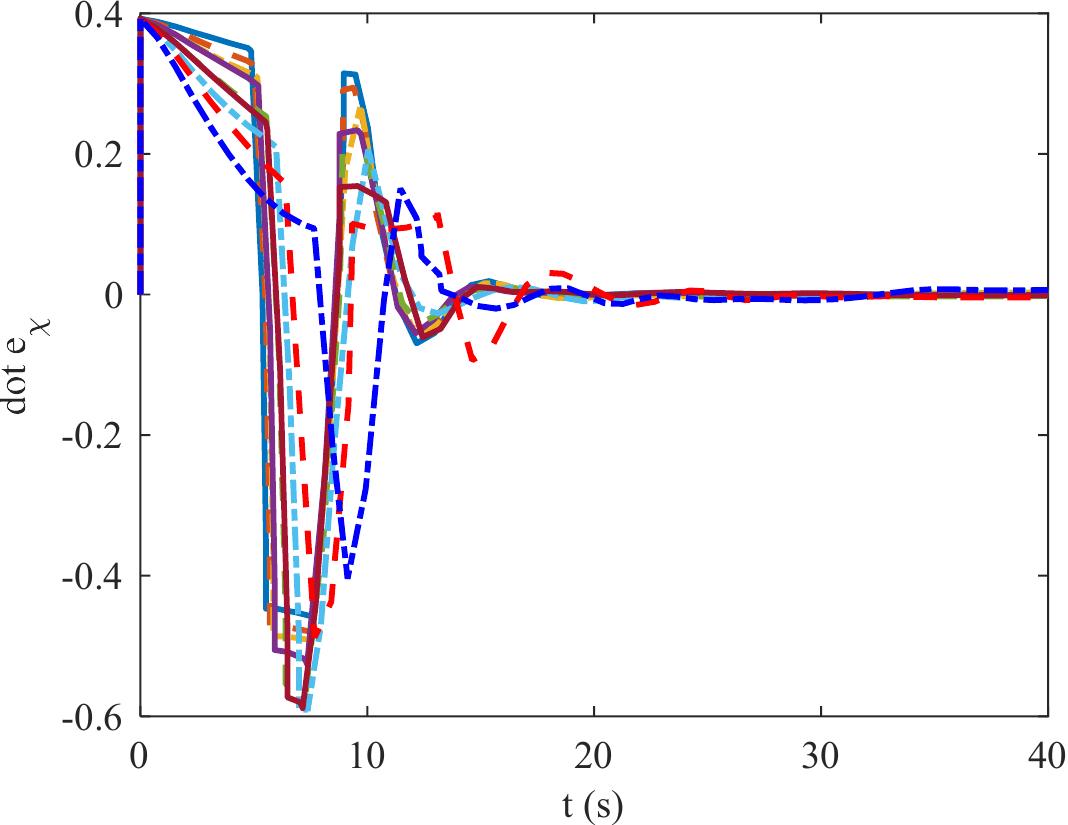} &
		\includegraphics[width=0.46\textwidth,height=0.85\columnwidth]{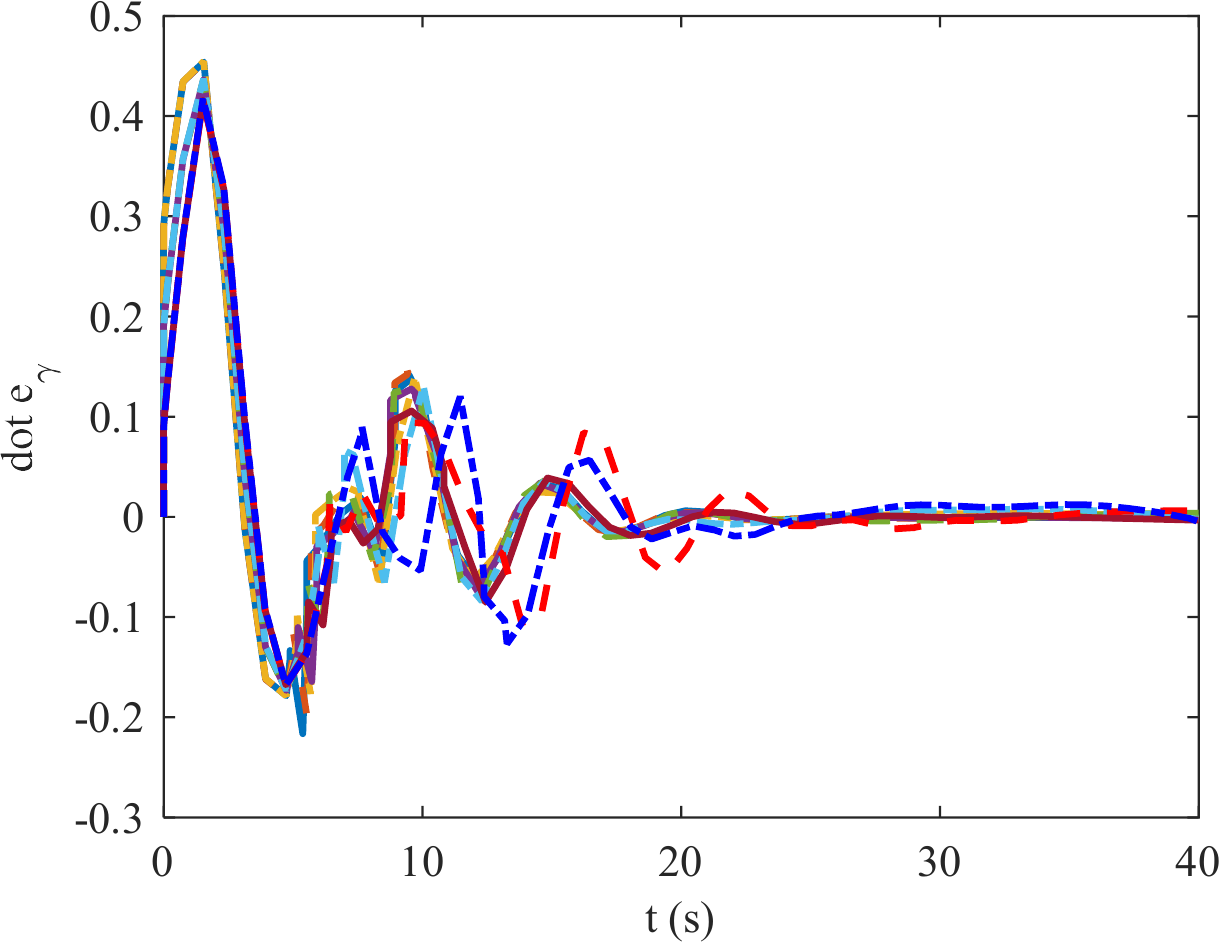} \\
		(c)  & (d) 
	\end{tabular}
	\caption{Error profiles comparsion of $\gamma$, $\chi$, $\dot{\gamma}$ and $\dot{\chi}$ under different $L_d$ and $\omega_d$.
		(a): Comparison of response profiles on $e_\chi$; 
		(b): Comparison of response profiles on $e_\gamma$;
		(c): Comparison of response profiles on $\dot{e}_{\chi}$;
		(d): Comparison of response profiles on $\dot{e}_{\gamma}$.
	}
	\label{Fig:11}
\end{figure*}

To evaluate the robust performance of the MIMO-PI controller under significant disturbances with different amplitudes and frequencies, we adopt the controller coefficients \( K_P^* \) and \( K_I^* \) and test its performance against disturbances characterized by various amplitudes \( L_d \) and frequencies \( \omega_d \) listed in Table \ref{tab:L_w} as follows
\begin{align}
	L_{d_{\chi}} = L_{d_{\gamma}} = L_d,\ \omega_{\chi} = \omega_{\gamma} = \omega_d 
\end{align}
The response curves of \( e_{\chi} \), \( e_{\gamma} \), \( \dot{e}_{\chi} \), and \( \dot{e}_{\gamma} \) under different disturbances are presented in Figure \ref{Fig:11}. It can be observed that nearly all response curves stabilize in the vicinity of the origin after 20 seconds. Notably, as disturbance intensity increases (i.e., with larger \( L_d \) and \( \omega_d \)), the overshoot of \( e_{\chi} \) decreases, while the post-20-second disturbance amplitudes of both \( e_{\chi} \) and \( e_{\gamma} \) increase accordingly. During this error stabilization process, despite the fact that larger \( L_d \) and \( \omega_d \) amplify the disturbance energy, the MIMO-PI controller effectively mitigates the impact of this energy, achieving robust stability near the origin to the greatest extent possible.
\begin{table}[htbp]
	\caption{The $L_d$ and $\omega_d$ for different disturbances.}
	\label{tab:L_w}
	%\centering
	\begin{tabular}{lll|lll|lll}
		\hline
		\bf{Type} & \bf{$L_d$}  & \bf{$\omega_d$} &
		\bf{Type} & \bf{$L_d$}  & \bf{$\omega_d$} &
		\bf{Type} & \bf{$L_d$}  & \bf{$\omega_d$}
		\\ 
		\hline % 横线
		$d_1$ & 0.1 & 0.1 &
		$d_2$ & 0.1 & 0.15 &
		$d_3$ & 0.1 & 0.2 \\
		$d_4$ & 0.2 & 0.1 &
		$d_5$ & 0.2 & 0.15 &
		$d_6$ & 0.2 & 0.2 \\
		$d_7$ & 0.3 & 0.1 &
		$d_8$ & 0.3 & 0.15 &
		$d_9$ & 0.3 & 0.2 \\
		\hline 
	\end{tabular}
\end{table}
\begin{figure*}[htbp]
	\centering
	\begin{tabular}{ccc}
		\includegraphics[width=0.32\textwidth]{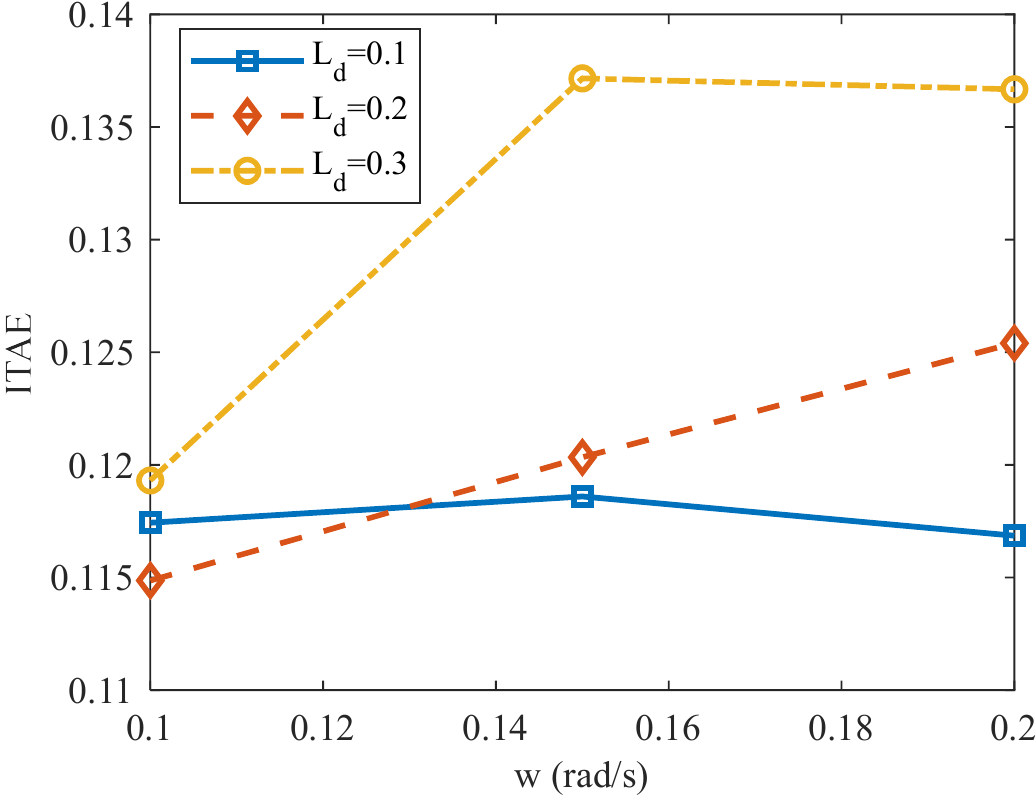} &
		\includegraphics[width=0.32\textwidth]{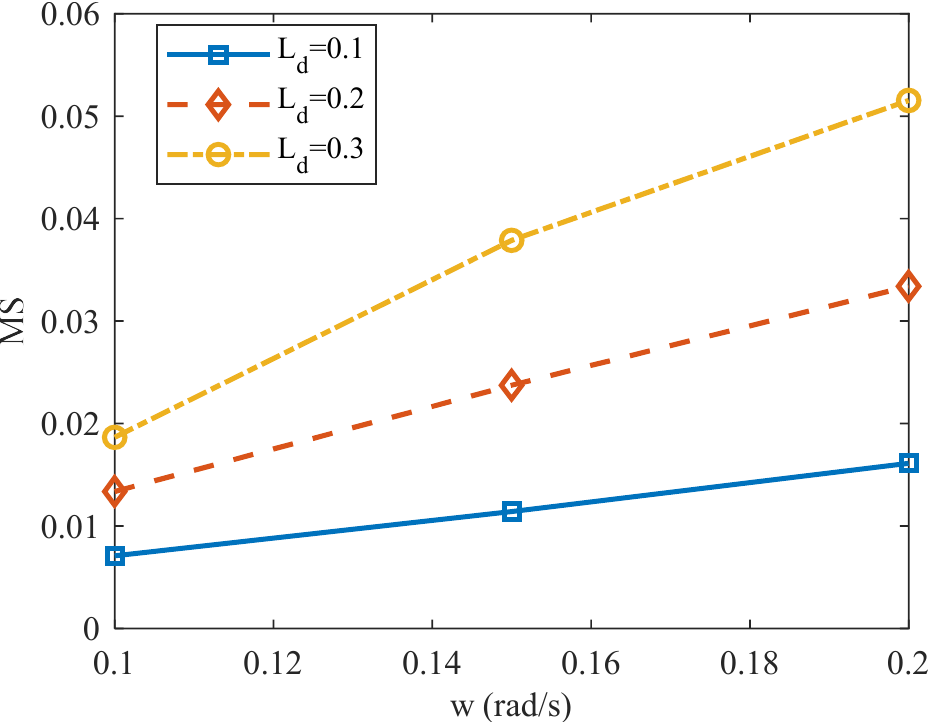} &	
		\includegraphics[width=0.31\textwidth]{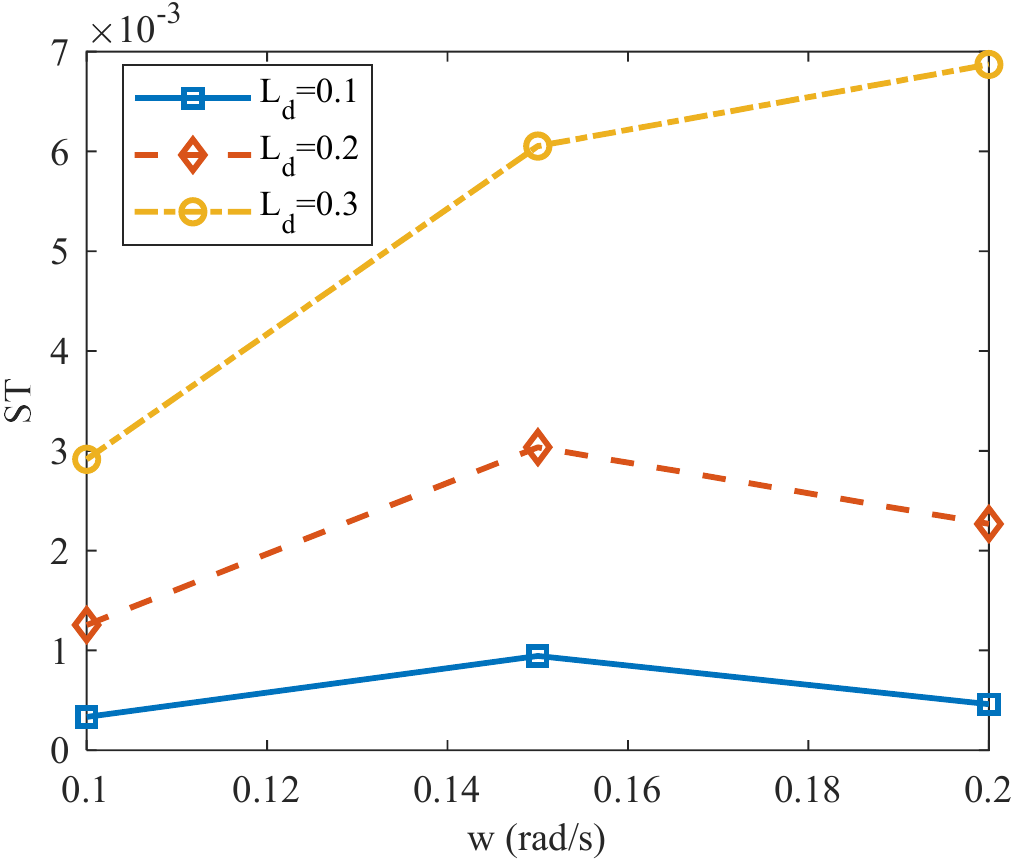}\\
		(a) & (b) & (c)\\
	\end{tabular}
	\caption{Comparison of ITAE, MS, and ST corresponding to different disturbance amplitudes \( L_d \) and frequencies \( \omega \). (a) Comparison of ITAE; (b) Comparison of MS; (c) Comparison of ST.}
	\label{Fig:10}
\end{figure*}

On the other hand, we quantified the ITAE, MS, and ST metrics under varying disturbances, as illustrated in Figure \ref{Fig:10}. First, analyzing from the ITAE perspective, it is observed that for disturbances with \( L_d = 0.1 \) and \( 0.3 \), increasing the disturbance frequency \( \omega_d \) does not result in a significant rise in ITAE; in fact, there is even a decreasing trend. This indicates that the MIMO-PI controller’s performance is insensitive to increases in disturbance frequency, reflecting its superior capability in suppressing high-frequency offsets.  Second, from the MS standpoint, increasing \( \omega_d \) and \( L_d \) enhances the disturbance-induced impact on steady-state performance. Fortunately, this impact exhibits linearity, which prevents excessively intense disturbances from triggering abrupt transitions or chaotic behaviors.  Finally, regarding ST: while an increase in \( L_d \) leads to a rise in ST, the magnitude of this steady-state standard deviation growth is minimal (on the order of \( 10^{-3} \)). Similarly, as \( \omega_d \) increases, ST shows no pronounced upward trend, indicating that the controller’s steady-state oscillation is also insensitive to \( \omega_d \) increments. This underscores its excellent performance in suppressing high-frequency oscillations.  

\section*{Conclusions and future work}
\label{sec:Conclusions and future work}
In this study, we put forward the critical quantitative metrics for assessing robustness in the context of general perturbed nonlinear system and validate its effectiveness. The specific research work is as follows: \\
(1) From the perspective of global random attractor, an innovative theoretical framework is proposed to quantify the stability performance of autonomous nonlinear disturbed systems;\\
(2) Building upon this theoretical framework, robust convergence indices \( R_K \) and \( I_K \) for quantifying MIMO-PI controllers are developed;\\
(3) A set of controller coefficient optimization methods is established to optimize the robust convergence indices. 

Experimental results validate the correctness of the proposed theory, the effectiveness of the indices, and the optimality of the controller.
Subsequent research should concentrate on two primary aspects: \\
(1) exploring methods to configure the eigenvalue distribution of $A_K(0)$ in order to attain an optimal robust indicator;\\
(2) investigating strategies to guarantee quadratic optimization of the error and input command through judicious parameter tuning while maintaining exponential convergence.
%\printcredits
%%% Loading bibliography style file,参考文献单开一页
%%\bibliographystyle{model1-num-names}
%%\bibliographystyle{cas-model2-names}
\bibliographystyle{unsrt}
%% Loading bibliography database
\bibliography{refs}
%\newpage
%\appendix
\end{document}